\documentclass[twocolumn,aps,prd,epsf]{revtex4-1}
\pdfoutput=1
\usepackage[pdftex]{graphicx}
\usepackage{color}
\usepackage{amsmath}
\usepackage{braket}
\usepackage[utf8]{inputenc}
\usepackage[T1]{fontenc}
\usepackage{url}
\usepackage[english]{babel}
\usepackage{graphicx}
\usepackage{caption}
\usepackage{subcaption}
\usepackage[export]{adjustbox}
\usepackage{dcolumn}
\usepackage{bm}
\usepackage{subfig}
\usepackage{float}
\usepackage{url}
\usepackage{verbatim}

\newcommand{\Tr}{\mbox{\rm Tr}}

\newcommand{\be}{\begin{equation}}
\newcommand{\ee}{\end{equation}}
\newcommand{\bea}{\begin{eqnarray}}
\newcommand{\eea}{\end{eqnarray}}
\newcommand{\non}{\nonumber}
\newcommand{\bi}{\begin{itemize}}
\newcommand{\ie}{\item}
\newcommand{\ei}{\end{itemize}}
\newcommand{\bn}{\begin{enumerate}}
\newcommand{\en}{\end{enumerate}}

\graphicspath{{figures/}}

\begin{document}

\title{ 
Tetraquark bound states and resonances in the unitary and microscopic triple string flip-flop quark model, the light-light-antiheavy-antiheavy $q q \bar Q\bar Q$ case study
}

\author{P. Bicudo}
\email{bicudo@tecnico.ulisboa.pt}
\author{M. Cardoso}
\email{mjdcc@cftp.ist.utl.pt}
\affiliation{CFTP, Dep. F\'{\i}sica, Instituto Superior T\'ecnico, Universidade de Lisboa, 
Av. Rovisco Pais, 1049-001 Lisboa, Portugal}

\begin{abstract}
We address $q q \bar Q\bar Q$ exotic tetraquark bound states and resonances with a fully unitarized and microscopic quark model.
We propose a triple string flip-flop potential, inspired in lattice QCD tetraquarks static potentials and fluxtubes,
combining meson-meson and tetraquark potentials. Our potential goes up to the color excited potential, but neglects spin-tensor potentials.
To search for bound states and resonances, we first solve the two-body mesonic problem. 
Then we develop fully unitary techniques to address the four-body tetraquark problem. We fold the four-body Shcr\"odinger equation with the mesonic wavefunctions, 
transforming it into a two-body meson-meson problem with non-local potentials. 
We find bound states for some quark masses numbers, including the one reported in lattice QCD.
Moreover, we also find resonances and calculate their masses and widths, by computing the $\mbox{T}$ matrix and finding it's
pole positions in the complex energy plane, for some quantum numbers.

However a detailed analysis of the quantum numbers where binding exists shows a discrepancy with recent lattice QCD results for the $l l \bar b \bar b$ tetraquark bound states. We conclude that the string flip-flop models need further improvement.
\end{abstract}
\maketitle


\section{Introduction \label{sec:intro}}

A long standing problem of QCD is the search for localized exotic 
states \cite{Jaffe:1976ig} and the corresponding decay to the hadron-hadron continuum. 
There is no QCD theorem preventing the existence of exotic hadrons, say two-gluon glueballs,
hybrids, tetraquarks, pentaquarks, three-gluon glueballs, hexaquarks, etc,
and the scientific community continues to search for clear exotic candidates.
However, this problem turned out to be much harder than expected.

In this work, we develop fully unitary techniques to solve some of the theoretical problems of multiquarks,
and predict multiquark bound states and resonances. We continue a previous unitary study of tetraquarks with a simplified two-variable toy model \cite{Bicudo:2010mv}, 
now fully solving the tetraquark problem, of the Schrödinger equation for two quarks and two antiquarks. 
We specialize in systems which are clearly exotic tetraquarks, who cannot have a significant mesonic quark-antiquark component, i e   where the quantum numbers, or the S-matrix pole and decay amplitudes clearly show it is a tetraquark.
 
\begin{figure}[t!]
\includegraphics[width=0.50\textwidth]{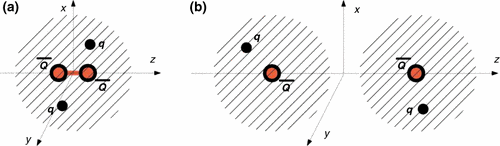}
\caption{Screening in the light-light -antistatic-antistatic tetraquark \cite{Bicudo:2012qt}.}
\label{fig:screening}
\end{figure}

In particular, as a benchmark, we study in detail the light-light-antiheavy-antiheavy systems who are expected to produce tetraquarks.
From basic principles of QCD, it is clear a system with two light quarks and two heavy antiquarks, for instance with flavour $u d \bar b \bar b$, should form a boundstate if the antiquarks are heavy enough \cite{Ader:1981db,Ballot:1983iv,Heller:1986bt,Carlson:1987hh,Lipkin:1986dw,Brink:1998as,Gelman:2002wf,Vijande:2003ki,Janc:2004qn,Cohen:2006jg,Vijande:2007ix}. To understand why binding should occur it is convenient to use the Born-Oppenheimer \cite{Born:1927} perspective, where the wavefunction of the two heavy antiquarks is determined considering an effective potential integrating the light quark coordinates. At very short $\bar b \bar b $ separations $r$, the $\bar{b}$ quarks interact with a perturbative one-gluon-exchange Coulomb potential, while at large separations the light quarks totally screen the interaction and the four quarks form two rather weakly interacting $B$ mesons as illustrated in Fig. \ref{fig:screening} . 
Thus a screened Coulomb potential is expected. This potential clearly produces a boundstate, providing the antiquarks $\bar b \bar b$ are heavy enough.

We leave the brief review of the tetraquarks in the literature for Section \ref{sec:review}, including the searches for tetraquarks in experiments, in dynamical lattice QCD and in quark models. We also discuss searches combining semi-dynamical lattice QCD with quantum mechanical techniques.
In Section \ref{sec:potential} we describe our triple string flip-flop potential,
where our system is open to the continuum in the inter meson-meson directions, and is confined in the tetraquark direction. Our string flip-flop potential also includes the first color excitation.
In Section \ref{sec:technique}, we address the meson-meson scattering problem, occurring when we solve the Schr\"odinger equation. We detail our numerical techniques, utilized to solve both our bound state and scattering problems. 
In Section \ref{sec:results} we show our results with the state of the art triple string flip-flop potential, exhibiting tetraquark bound states, and resonances. For the resonances, with our fully unitarized computation, we calculate the pole position and thus finding their decay widths. We also consider, to compare with our full computation, simplified potentials. 
In Section \ref{sec:conclu} we compare our results to the existing lattice QCD results for the light-light-antiheavy-antiheavy system and conclude. We find excessive binding, concluding the flip-fop potentials need further improvement. Nevertheless our technique can be applied to other potentials who may be developed in the future.


\section{Brief review of tetraquarks \label{sec:review}}

\subsection{The experimental search for tetraquarks}

This is a very difficult problem experimentally, since exotic candidates are resonances immersed in the excited hadron spectra, and moreover they usually decay to several hadrons.

Recently an experimental article \cite{Alekseev:2009aa} was published indicating that the $\pi_1(1600)$  
observation of a resonance with  mass $M= 1660 (74)$ MeV , and width $\Gamma$= 269 (85) MeV, in diffractive dissociation of 190 GeV/c $\pi -$ into $\pi^- \,  \pi^- \, \pi^+$,
has exotic $J^{PC} = 1^{-+}$ quantum numbers, consistent with an hybrid meson or a tetraquark. 

Moreover, the existence of tetraquarks has been advanced by the experimental collaborations at the charm and bottom factories to interpret the new $X, \, Y, \, Z$ hidden charm or bottom mesons 
\cite{Godfrey:2008nc}. 
Their mass and decay products mark them as charmonia-like resonances but their masses do not fit into the quark model spectrum of quark-antiquark mesons 
\cite{Godfrey:1985xj}. This class of tetraquarks is related to the double-heavy tetraquarks we address here.

In particular, the charged $Z_c^{\pm}$ and $Z_b^{\pm}$ are crypto-exotic, but technically they can be regarded as essentially exotic tetraquarks if we neglect $ c \bar c$ or $b \bar b$ annihilation.
There are two $Z_b^{\pm}$ observed only by the collaboration BELLE at KEK
\cite{Belle:2011aa}, 
slightly above $B \ \bar B^*$ and $B^* \ \bar B^*$ thresholds,
 the $Z_b(10610)^+$ and $Z_b(10650)^+$.
Their nature is possibly different from the  two $Z_c(3940)^{\pm}$and  $Z_c(4430)^{\pm}$, whose mass is well above $DD$ threshold \cite{Choi:2007wga}. The $Z_c^\pm$ has received a series of experimental observations by the BELLE collaboration \cite{Liu:2013dau,Chilikin:2014bkk}, the Cleo-C collaboration \cite{Xiao:2013iha}, the
BESIII collaboration \cite{Ablikim:2013mio,Ablikim:2013emm,Ablikim:2013wzq,Ablikim:2013xfr,Ablikim:2014dxl} and the LHCb
collaboration \cite{Aaij:2014jqa}. This family is possibly related to the closed-charm pentaquark recently observed at LHCb \cite{Aaij:2015tga}.

However, to establish a new resonance it is necessary to study with an accurate level of confidence all its properties, including its mass and width as determined by its $S$-matrix pole and all relevant partial decay widths. 
Possibly we need more data and more extensive data analysis to be able to absolutely confirm exotics
\cite{Dzierba:2003cm,Dzierba:2005jg}.

Notice, using very approximate Resonant Group Method calculations, in 2008, we predicted
\cite{Cardoso:2008dd}
a partial decay width to $\pi \ J/\psi $ of the $Z_c(4430)^-$ consistent with the experimental value.
But we opt to leave the detailed study of this second class of double-heavy tetraquarks, with closed charm or closed bottom tetraquarks, for future studies. We first want to address more constrained  systems with identical fermions, where the Pauli symmetry applies.

\subsection{The lattice QCD search for tetraquarks}

\begin{figure}[t!]
\begin{center}
\includegraphics[width=1.\linewidth]{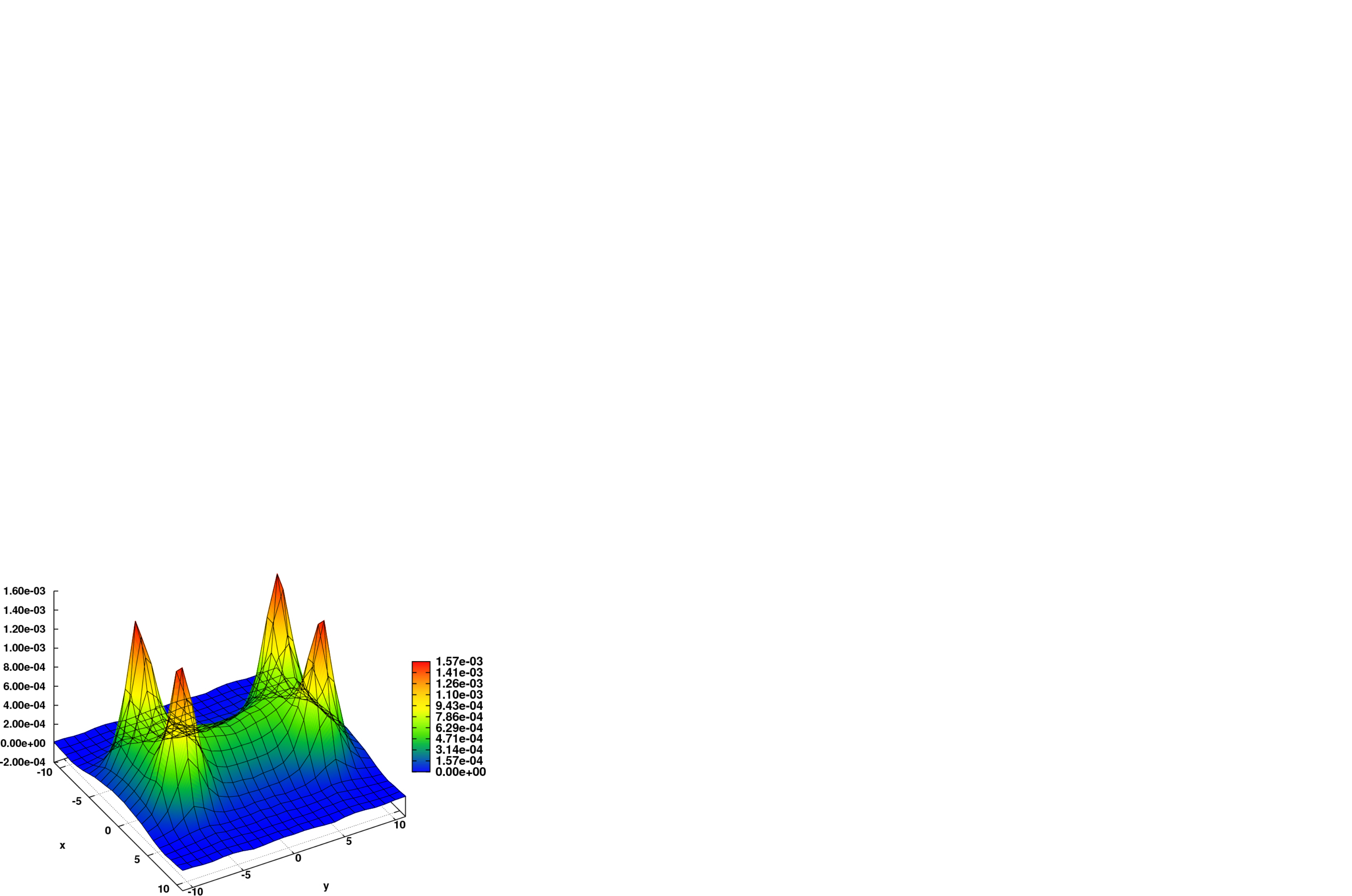}
  \caption{Surface plot of the static tetraquark flux tube computed in lattice QCD
 \cite{Cardoso:2011fq,Cardoso:2012uka} . 
\label{TQ_EB_ape_hyp_r1_8_r2_14_Act_3D_Sim}
}
\end{center}
\end{figure}

In lattice QCD, the study of exotics is presently even harder than in the laboratory, 
since the techniques and computer facilities necessary to study of resonances with many decay channels remain under development.

Thus lattice QCD started by searching for the expected boundstate in light-light-antiheavy-antiheavy channels \cite{Ikeda:2013vwa,Guerrieri:2014nxa}. Using dynamical quarks, the only heavy quark presently accessible to Lattice QCD simulations is the charm quark. No evidence for boundstates in this family of tetraquarks, say for a $u d \bar b \bar b$ was found.

Lattice QCD also searched for evidence of a large tetraquark component in closed bottom the $Z_c(3940)^-$ candidate 
\cite{Prelovsek:2014swa,Leskovec:2014gxa}.
The difficulty of the study of the $Z_c^-$, a resonance well above threshold, is due to its many two-meson coupled channels. 
The authors considered 22 two-meson channels, corresponding to lattice QCD interpolators $O^{M_1 \, M_2}$. In addition they considered 4 tetraquarks channels, corresponding to  diquark-antidiquark interpolators with flavour and color $  [\bar c \bar u]_3  \ [c d]_{\bar 3}$.
An evidence for the tetraquark resonance candidate was investigated in the full coupled correlator matrix of hadron operators. 

Finally, after switching on and off the 4 tetraquark-like channels, the authors \cite{Prelovsek:2014swa,Leskovec:2014gxa} found no significant deviation in the 13 lowest channels, who span the energy range from the lowest threshold to the $Z_c(3940)^-$ candidate. Thus, the authors concluded there is no robust evidence of a $Z_c^{\pm}$ tetraquark resonance.

However, the direct proof for, or against, a tetraquark resonance in lattice QCD would require the study of the S-matrix. The technique to perform phase shift analysis in lattice QCD exists
\cite{Luscher:1986pf,Luscher:1985dn}. 
From the phase shift analysis, inasmuch as with experimental data, the poles of the  S-matrix can be extracted. But phase shift analysis of the tetraquark $Z_c^+$ has not yet been done with lattice QCD data. For an absolute evidence, the different partial decay widths should be computed as well in lattice QCD.

With present computers only resonances with just $\sim$ 1 open decay channel have been studied in lattice QCD, with sufficient detail
\cite{Lang:2011mn}.
The method of extracting the phase shifts from the spectrum of harmonic waves in a box has been extended to inelastic (more than one open channel) coupled channels 
\cite{Dudek:2014qha}. 
However, tetraquarks are excited resonances, decaying into many channels, $\sim$ 30 for the last experimental $Z_c^-$ candidates.
This is presently unattainable by present computers and codes, but we expect Lattice QCD to be on the way to, in the future, reach the level of experimental data analysis.

\subsection{Quark model approaches to exotic tetraquarks}

A detailed theoretical understanding of the properties of exotic hadrons is important to support the experimental and lattice QCD searches of exotics. 
Several theoretical problems need to be solved before addressing exotics hadrons.

\subsubsection{Early quark models }

 Already at the onset of QCD, the bag model predicted many tetraquarks
 \cite{Jaffe:1976ig}. 
However, as soon as lattice QCD was able to compute static quark potentials and color electromagnetic fields , it was realized that quark confinement was not bag-like, but string-like, due to color flux tubes. 
 
Inspired in lattice QCD linear confining potentials,   the relativized quark model potential was developed  
\cite{Godfrey:1985xj, Capstick:1986bm}, after the authors fitted the spectra of  all known hadrons in the 80's. 
Notice that a correctly calibrated quark model needs many terms and many parameters, say of the order of $\sim$ 20 parameters.

Nevertheless the relativized quark model still lacks two crucial effects,
leading with up to 400 MeV deviations,
chiral symmetry breaking, which has already been for instance in the Dyson-Schwinger approach
\cite{Bicudo:1989sh,Bicudo:1989si,Bicudo:1989sj},
and coupled channels / unquenching which has already been include for instance in effective meson or baryon models
\cite{Rupp:2012py}.

\subsubsection{Extra binding with four-body flux tube potentials}

\begin{figure}[t!]
\hspace{-50pt}
\includegraphics[width=1.3\linewidth,left]{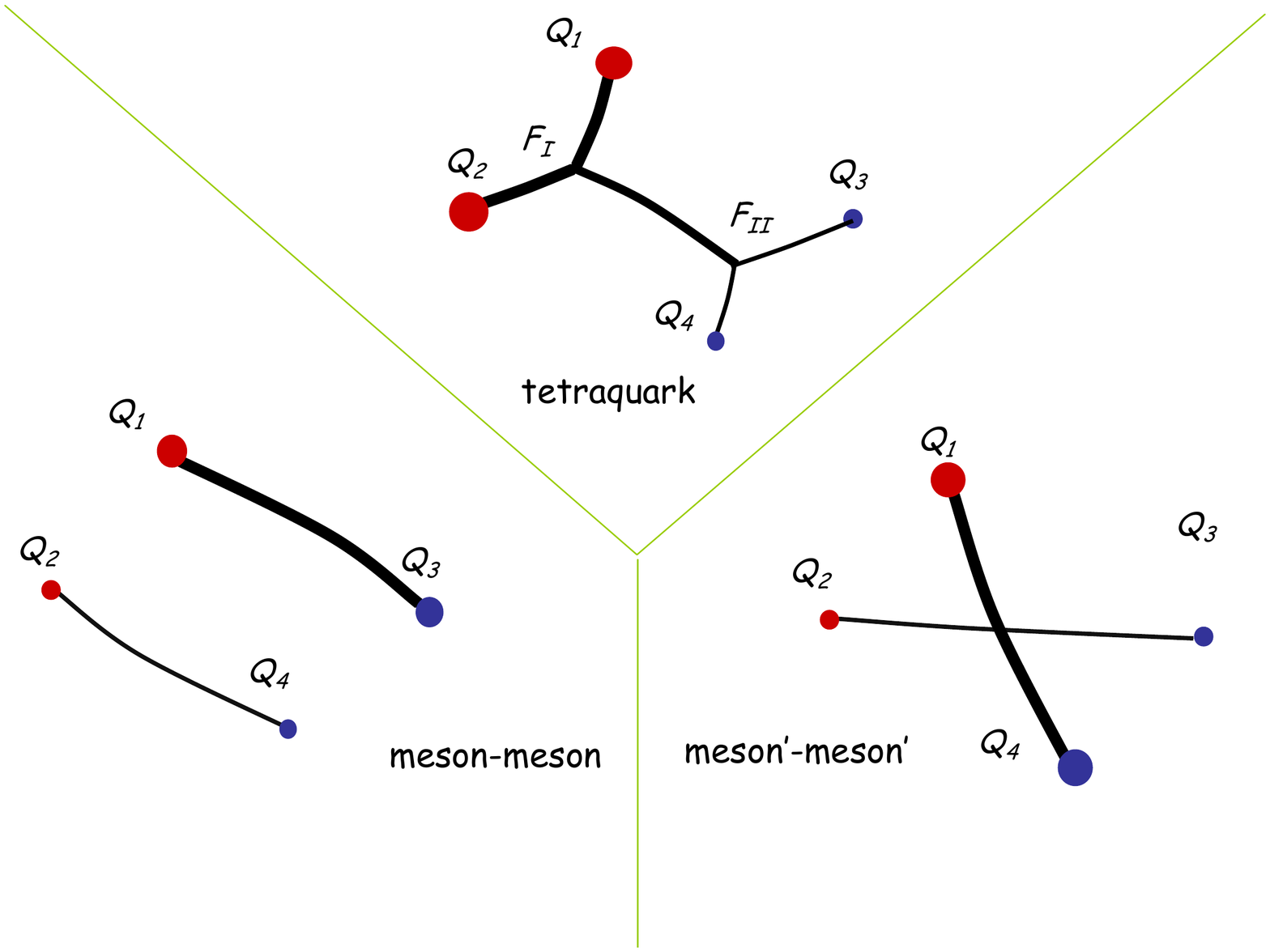}
  \caption{Triple string flip-flop Potential potential. While the previous string flip-flop potentials choose the minimum of two different meson pair potentials, we consider as well the tetraquark potential \cite{Bicudo:2010mv}. 
\label{tripleflipflop}
}
\end{figure}

Moreover, since tetraquarks are always open to decays into a meson-meson pair, tetraquark resonances or bound states may only exist if a mechanism exists to provide binding specific to tetraquarks.
Here we explore a mechanism observed in lattice QCD static potentials: the confining four body potential \cite{Alexandrou:2004ak,Okiharu:2004ve},  produced by double Y or butterfly shaped flux tubes or strings \cite{Cardoso:2011fq,Cardoso:2012uka,Cardoso:2012rb}, depicted in Fig. \ref{TQ_EB_ape_hyp_r1_8_r2_14_Act_3D_Sim}. This mechanism is related to the Jaffe-Wilczek model 
\cite{Karliner:2003dt,Jaffe:2004zg,Jaffe:2004wv}
who proposed the tetraquark would form a diquark-antidiquark system. 

We acknowledge other mechanisms may exist to support binding, however they are complex to implement and it is not clear, from lattice QCD, how they work quantitatively. 
For instance attraction may also be due to quark-antiquark annihilation, however it turned out to be insufficient to bind a proposed pentaquark \cite{Bicudo:2003rw,LlanesEstrada:2003us}. Another mechanism is the hyperfine spin-dependent potential utilized in the original bag model \cite{Jaffe:1976ig}, however the spin-tensor potentials have only been computed in lattice QCD for mesons. For baryons they are model dependent, while for in multiquarks the details of the spin-tensor potentials remain speculative.  Moreover both quark-antiquark annihilation and the hyperfine potential are also important for chiral symmetry breaking. To avoid the complexity of quark-antiquark annihilation, spin tensor quark-quark interactions and chiral symmetry breaking, we specialize in a family of tetraquarks where they can be neglected.

Here we consider purely exotic tetraquarks only. In contradistinction to crypto-exotics, in pure exotics quark-antiquark annihilation does not occur directly. Crypto-exotic tetraquarks are also less clear in the sense they always have a mesonic component, they are never a pure exotic. Thus we consider only tetraquarks with exotic flavour. Moreover, as a first study, we neglect spin-tensor interactions and chiral symmetry breaking effects.

\subsubsection{Solving the Van der Waals force problem with string flip-flop potentials}

Clearly, tetraquarks are always coupled to meson-meson systems, and we must be able to address correctly the meson-meson interactions.

Notice confining two-body potentials with the SU(3) color Casimir invariant 
$ \vec \lambda_i \cdot \vec \lambda_j $ suggested by the One-Gluon-Exchange potential,
and possibly compatible with lattcie QCD, 
lead to a Van der Waals potential,
\be
V_\text{Van der Waals}  = { V'(r) \over r}
 \times T.
\ee
where $T$ is a polarization tensor.
This would lead to an extremely large Van der Waals 
\cite{Fishbane:1977ay,Appelquist:1978rt,Willey:1978fm,Matsuyama:1978hf,Gavela:1979zu,Feinberg:1983zz}
force between mesons, or baryons which clearly is not observed experimentally.
Thus two-body confinement dominance is ruled out for multiquark systems.

The string flip-flop potential for the meson-meson interaction was developed
\cite{Miyazawa:1979vx,Miyazawa:1980ft,Oka:1984yx,Oka:1985vg,Karliner:2003dt},
to solve the problem of the Van der Waals forces produced by the two-body
confining potentials.

Traditionally, the string flip-flop potential considers that the potential is the one
that minimizes the energy of the possible two different meson-meson configurations, say $M_{13} \, M_{24}$ or $M_{14} \, M_{23}$.
This removes the inter-meson potential, and thus solves the problem of the Van der Waals force.

Here we upgrade the string flip-flop potential, considering a third possible
configuration, the tetraquark one, say $T_{12 \, , \, 34}$,
where the four constituents are linked
by a connected string
\cite{Vijande:2007ix,Vijande:2009xx}. 
We study whether the tetraquark attractive flux tube may induce further binding of tetraquarks.

The three configurations differ in the strings linking the quarks and antiquarks, 
this is illustrated in 
Fig. \ref{tripleflipflop}. 
When the diquarks $qq$ and $\bar q \bar q$ have small distances, the tetraquark configuration minimizes the string energy.
When the quark-antiquark pairs $q \bar q$ and $q \bar q$ have small distances, the meson-meson configuration minimizes
the string energy.

\subsubsection{Previous quark model results with string flip-flop potentials}

Tetraquark resonances are quite subtle, since the Fock space of tetraquarks is the same of its decay channels, the meson-meson channels, and moreover a potential barrier is absent. A priori it is not intuitive whether this system may produce resonances or not. 

Nevertheless, there is an argument \cite{Karliner:2003dt} suggesting multiquarks with angular excitations may gain a centrifugal barrier, leading to narrower decay widths.

With a triple string flip-flop potential, 
bound states below the threshold for hadronic coupled channels have been found
\cite{Beinker:1995qe,Zouzou:1986qh,Vijande:2007ix,Vijande:2009xx}.

On the other hand, the string flip-flip potentials allow fully unitarized studies of resonances
\cite{Lenz:1985jk,Oka:1984yx,Oka:1985vg}.
Utilizing analytical calculations with a double flip-flop harmonic oscillator potential
\cite{Lenz:1985jk},
and using the resonating group method again with a double flip-flop confining harmonic oscillator potential
\cite{Oka:1984yx,Oka:1985vg},
resonances and bound states have already been predicted.

Moreover, using the perturbative approximation of the resonating group method, a preliminary estimation of the partial decay width of the $Z(4430)^-$ resonance was similar to the one measured by LHCb
\cite{Cardoso:2008dd}. 

These studies suggest tetraquark bound states or resonances are plausible. Fully unitarized techniques adapted to state of the art potentials, remain to be applied in order to achieve a quantitative theoretical study of tetraquark resonances and bound states.

\begin{figure}[t!]
\includegraphics[width=0.50\textwidth]{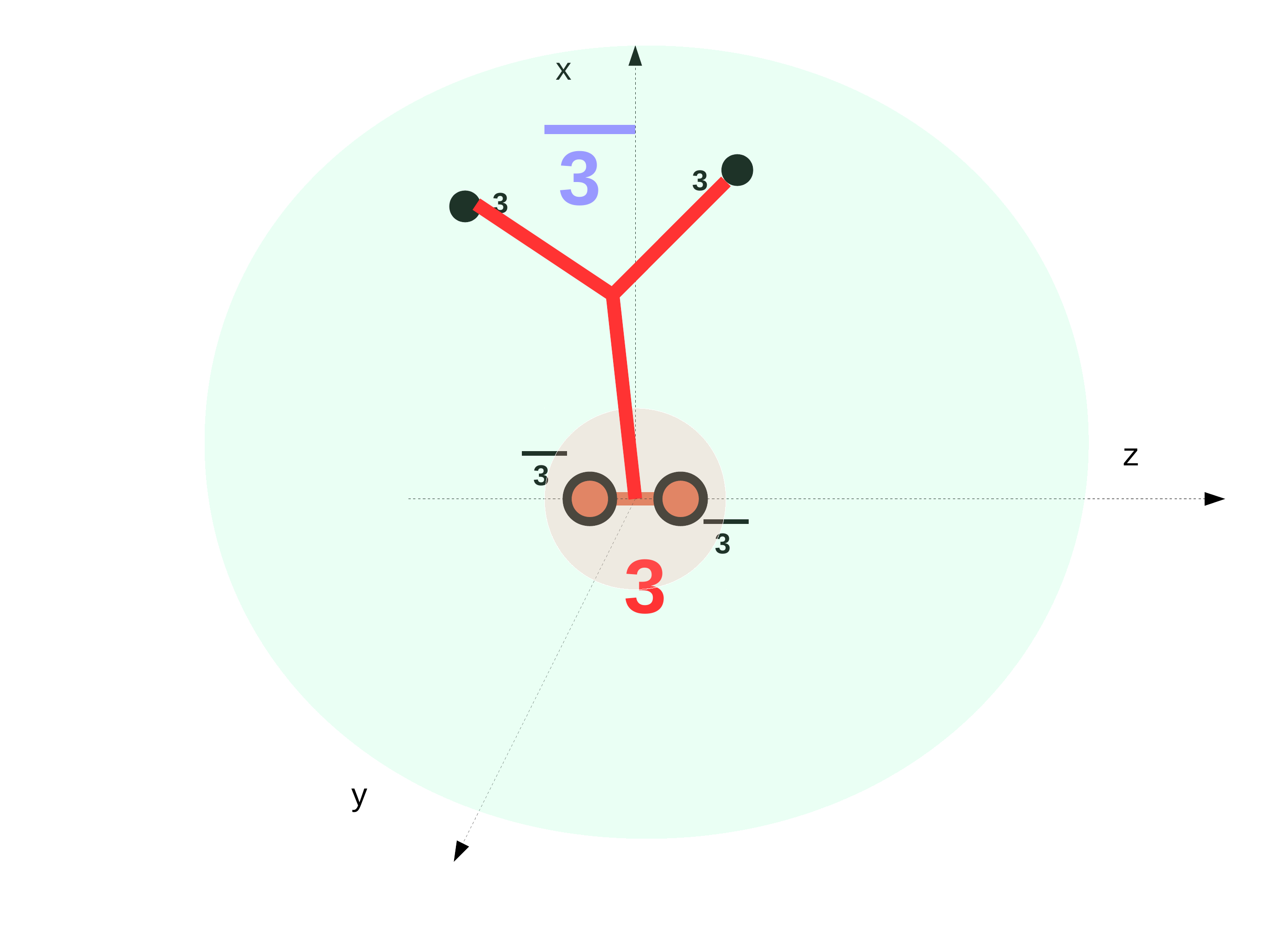}
\caption{Color of the diquarks determined by the quantum numbers of $BB$ tetraquark attractive channels \cite{Bicudo:2015vta}.}
\label{fig:quantumnumbers}
\end{figure}

\subsection{Recent lattice QCD results for the potentials in the light-light-antistatic-antistatic system.}

The difficulties of both lattice QCD and quark models can be both relaxed when one uses a hybrid approach with both numerical and model techniques.

\begin{figure}[t!]
\includegraphics[width=0.50\textwidth]{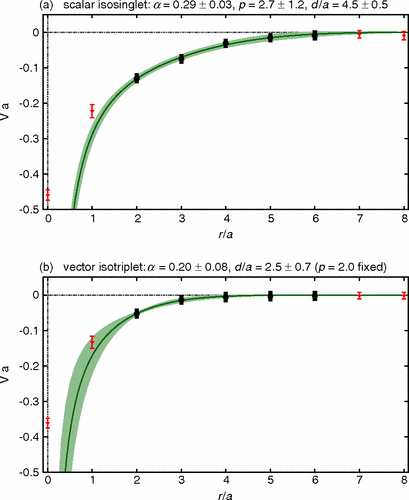}
\caption{Attractive channels in lattice QCD \cite{Bicudo:2012qt}.}
\label{fig:attraction}
\end{figure}

Recently, the light-antistatic light-antistatic potentials have been computed in in lattice QCD \cite{Wagner:2010ad, Wagner:2011ev}, as shown in Fig. \ref{fig:attraction} .  A static antiquark constitutes a good approximation to a spin-averaged $\bar b$ bottom antiquark. The light-antistatic light-antistatic potential can then be used, with the Born-Oppenheimer approximation \cite{Born:1927}, as the a  $B-B$ potential,
where we the higher order $1/m_b$ terms including the spin-tensor terms are neglected. According to the quantum numbers of the two dynamical light quarks, either attraction or repulsion is found. 
The attraction/repulsion can be understood with the screening mechanism illustrated in Fig. \ref{fig:screening} , and the antistatic-antistatic potential is well fitted by the screened Coulomb ansatz \cite{Bicudo:2012qt},
\be
V(r)=-\frac{\alpha}{r}\exp\left(-\left(\frac{r}{d}\right)^p\right) .
\label{eq:ansatzBB}
\ee

\begin{figure}[t!]
\includegraphics[width=0.50\textwidth]{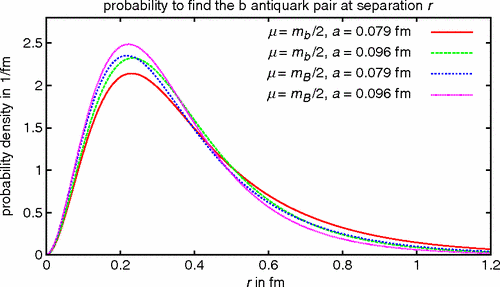}
\caption{Probability density of the $\bar b \bar b$ pair determined with the Born Oppenheimer approximation \cite{Bicudo:2012qt}.}
\label{fig:latBOWF}
\end{figure}

Utilizing the potential of the channel with larger attraction, occurring in the Isospin=0 and Spin=0 quark-quark system, together with the Born-Oppenheimer approximation  \cite{Born:1927} to derive the Schr\"odinger equation, 
the possible boundstates of the heavy antiquarks are then investigated with quantum mechanics techniques.
Recently, this approach indeed found evidence for a tetraquark $u d \bar b \bar b$ boundstate \cite{Bicudo:2012qt,Brown:2012tm}, while no boundstates have been found  for states where the heavy quarks are $\bar c \bar b$ or $\bar c \bar c$ (consistent with full lattice QCD computations \cite{Ikeda:2013vwa,Guerrieri:2014nxa}) or where the the light quarks are $\bar s \bar s$ or $\bar c \bar c$ \cite{Bicudo:2015vta}. The $\bar b \bar b$ probability density in the only binding channel is shown in Fig. \ref{fig:latBOWF}.

Moreover, lattice QCD finds attraction or repulsion, consistently with the screening model.
\bi
\ie In some channels we find attraction, in others we find repulsion. 
\ie Phenomenologically, due to the short range one-gluon exchange proportional to the Gell-Mann $ \lambda^i \cdot \lambda^j$ Casimir operator, the $\bar b$ and $\bar b$ are expected to bind only if they are in a triplet 3, not a sextet 6,
\ie thus the light quarks $q q $ should be in a color antitriplet $\bar 3$, as illustrated in Fig. \ref{fig:quantumnumbers},
\ie due to the Pauli principle and to the quark model phenomenology this is indeed consistent with  an s-wave, I=0 and S=0 diquark.
\ei
These results for repulsion, attraction or binding are very important, the quark models should comply with them.

\section{Potential \label{sec:potential}}

\subsection{Groundstate string flip-flop potential}

\begin{figure*}[t!]
        \centering
        \begin{subfigure}[b]{0.45\textwidth}
                \includegraphics[trim=2.1cm 0.4cm 1.1cm 0.6cm, clip,width=\textwidth]{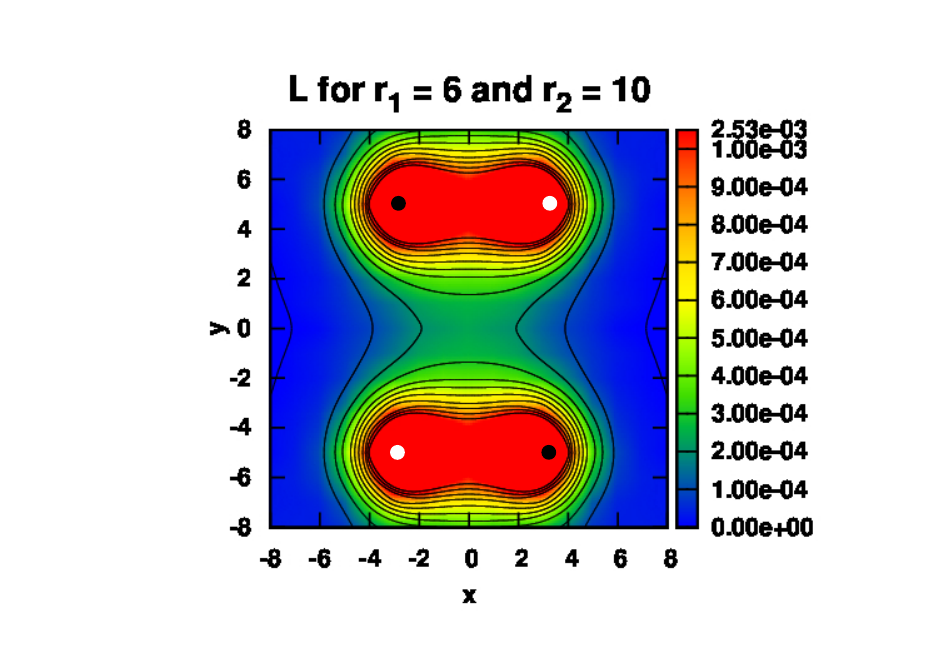}
                \caption{Meson-meson groundstate flux tube}
                \label{fig:meson}
        \end{subfigure}
        ~ 
          \qquad
        \begin{subfigure}[b]{0.45\textwidth}
                \includegraphics[trim=2.1cm 0.4cm 1.1cm 0.6cm, clip,width=\textwidth]{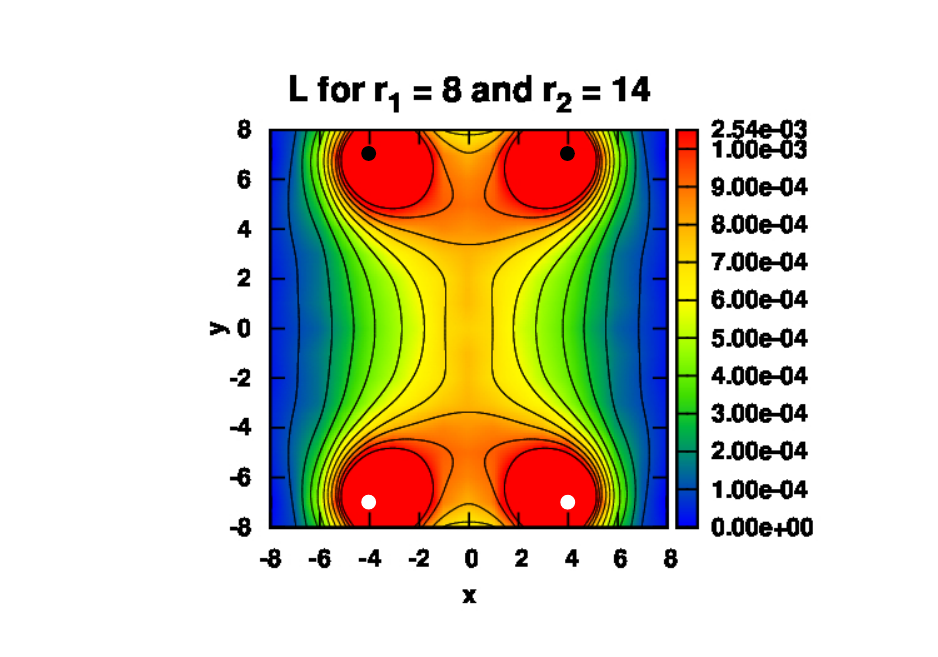}
                \caption{Tetraquark groundstate flux tube}
                \label{fig:tetraq}
        \end{subfigure}
        \caption{Lagrangian density plots for static quark-quark-antiquark-antiquark flux groundstate tubes computed in lattice QCD \cite{Cardoso:2012uka}. The density and the coordinates are in lattice spacing units \cite{Cardoso:2012uka}. Three different flux tubes occur for different geometries of the four-body system.}
        \label{fig:tetraqmeson1meson2}
\end{figure*}

\begin{figure*}[t!]
        \centering
        \begin{subfigure}[b]{0.45\textwidth}
                \includegraphics[trim=2.1cm 0.4cm 1.1cm 0.6cm, clip,width=\textwidth]{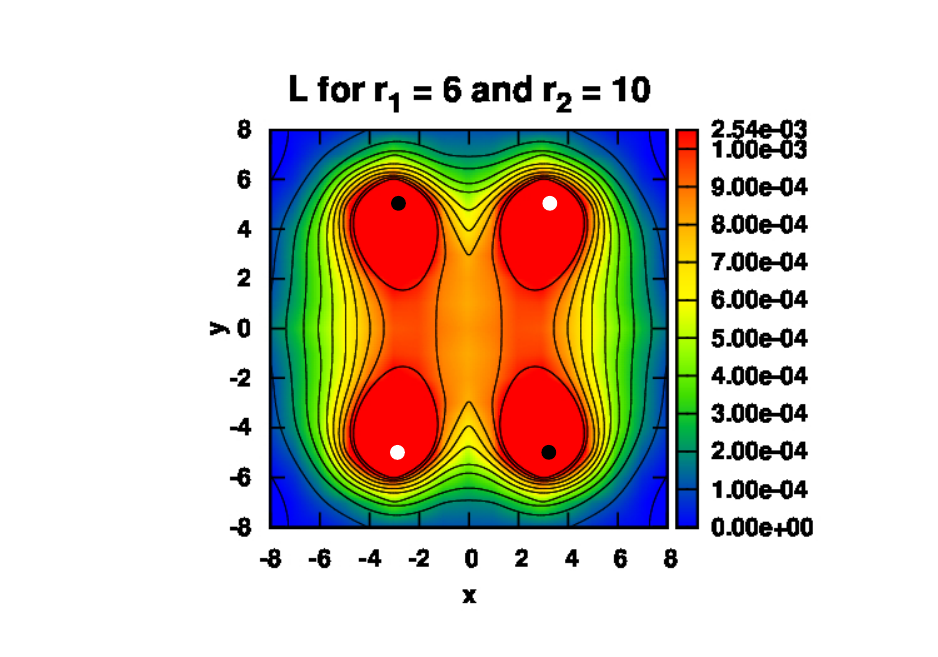}
                \caption{Meson-meson excited flux tube}
                \label{fig:mesonex}
        \end{subfigure}
        ~ 
          \qquad
        \begin{subfigure}[b]{0.45\textwidth}
                \includegraphics[trim=2.1cm 0.4cm 1.1cm 0.6cm, clip,width=\textwidth]{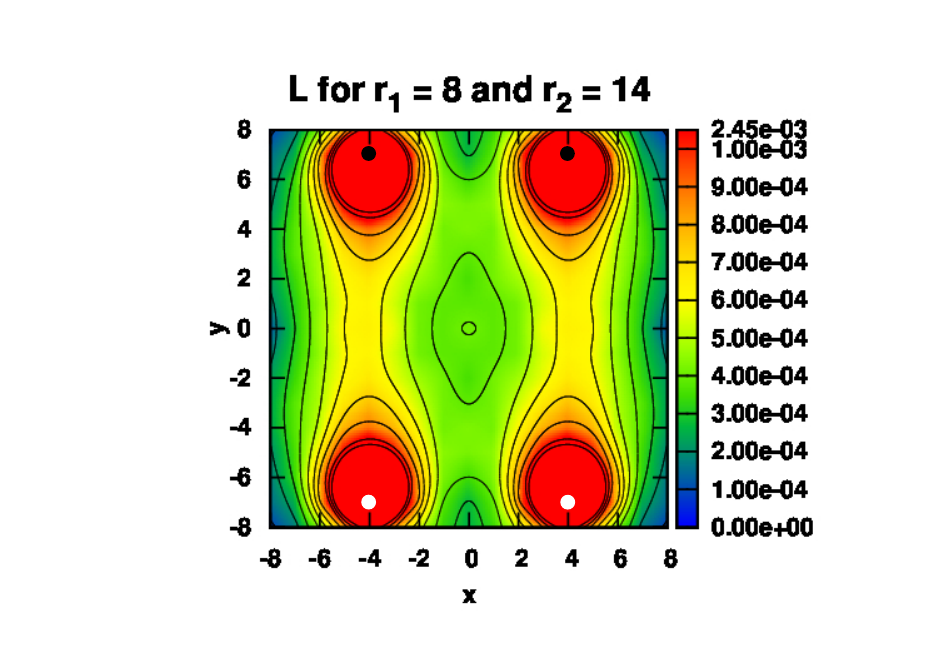}
                \caption{Tetraquark excited flux tube}
                \label{fig:tetraqex}
        \end{subfigure}%
        \caption{Lagrangian density plots for static quark-quark-antiquark-antiquark first excitation flux tubes computed in lattice QCD \cite{Cardoso:2012uka}. The density and the coordinates are in lattice spacing units \cite{Cardoso:2012uka}. We show the first excitation of the different groundstate flux tubes shown in Fig.  \ref{fig:tetraqmeson1meson2}. In \ref{fig:mesonex} the excitation of a meson-meson flux tube is similar to the other meson-meson flux tube. In \ref{fig:tetraqex} the excitation of a tetraquark flux tube is similar to a meson-meson flux tube.
        }
        \label{fig:tetraqmeson1meson2ex}
\end{figure*}

We know from lattice computations for static quarks \cite{,Alexandrou:2004ak,Okiharu:2004ve,Cardoso:2011fq,Cardoso:2012uka} that the
ground state potential for a system composed of two quarks and two
antiquarks, is well fitted by a string flip-flop potential ,
\be
V^0_{FF}=\min(V_{M\hspace{-1.2pt}M},V_{M\hspace{-1.2pt}M'},V_{T}) \ ,
\label{eq:3flip-flop}
\ee
were, for simplicity, we neglect possible mixings at the transition
regions. In Eq. (\ref{eq:3flip-flop}), $V_{M\hspace{-1.2pt}M}$ and $V_{M\hspace{-1.2pt}M'}$ are two possible potentials of a pair of
mesons, given by the sum of the intra-meson potentials $V_{M\hspace{-1.2pt}M}(\mathbf{x}_{i})=V_{M}(|\mathbf{x}_{1}-\mathbf{x}_{3}|)+V_{M}(|\mathbf{x}_{2}-\mathbf{x}_{4}|)$
and $V_{M\hspace{-1.2pt}M'}(\mathbf{x}_{i})=V_{M}(|\mathbf{x}_{1}-\mathbf{x}_{4}|)+V_{M}(|\mathbf{x}_{2}-\mathbf{x}_{3}|)$.
The intra-meson potential is well described by the funnel potential, 
\be
V_{M}(\mathbf{r}_{ij})=K-\frac{\gamma}{r_{ij}}+\sigma r_{ij} \ ,
\ee
where $r_{ij}=|\mathbf{x}_{i}-\mathbf{x}_{j}|$ and 
which includes a constant term $K$, the short range Coulomb potential $-\frac{\gamma}{r}$ and the long range confining potential $\sigma r$. Here we use the indices $1$ and $2$, to refer to the quarks while
$3$ and $4$, refer to the antiquarks. $\mathbf{x}_{i}$ are the
positions of the quarks/antiquarks. Moreover the tetraquark function $V_{T}$ is, 
\be
V_{T}(\mathbf{x}_{i})=2K-\sum_{1=i<j}^4C_{ij}\frac{\gamma}{r_{ij}}+\sigma L_{min}(\mathbf{x}_{i}) \ ,
\ee
where $C_{ij}=1/2$ for quark-quark and antiquark-antiquark and $C_{ij}=1/4$
for quark-antiquark interactions.
$L_{min}$ is the minimal distance linking the four particles,
\be
L_{min}(\mathbf{x}_{i}) =  r_{1\, 5} + r_{2\, 5} + r_{5\, 6} + r_{3\, 6} + r_{4\, 6} \ ,
\ee
where $5$ and $6$ are the indices of the two Fermat-Torricelli-Steiner points 
\cite{Vijande:2007ix,Vijande:2009xx,Bicudo:2008yr} of the tetraquark. We compute the position of these two points with the numerical technique of Ref. \cite{Bicudo:2008yr}.
This potential generalizes the earlier string flip-flop potential
models by introducing a third possible branch in the potential where
the four particles are all linked by a confining string. 

Our main approximation consists in neglecting the spin tensor potentials, say the hyperfine, spin-orbit and tensor potentials. These potentials have only been studied in detail in the two-body mesonic quark-antiquark system, including in lattice QCD computations \cite{Koma:2006fw}. 
In the three quark baryonic system, the spin tensor potentials have been applied in models, however they remain to be computed in lattice QCD. In four-body tetraquarks systems, we still ignore the spin-tensor potentials.  Moreover, the study of tetraquarks with a four-body confining potential remains an open problem, and our present aim is to solve it before addressing quark spin interactions. Thus our tetraquark masses or poles should be seen as spin-averaged results.

\subsection{Color states}

However, this potential is not yet sufficiently defined to completely address tetraquarks.
It has been pointed out 
\cite{Green:1994cg,Green:1995df,Pennanen:1998nu,Bornyakov:2005kn,Akbar:2011nh}
that we also need a potential for color excited states,
otherwise there would be solutions which wouldn't be color singlets. 

With two quarks and two antiquarks, two different linearly independent color
singlets can be constructed. Physically, rather than choosing a basis composed by singlet-singlet and octet-octet, or including the triplet-antitriplet, it is more convenient to have a color basis corresponding to the two possible asymptotic meson-meson systems $MM$ and $MM'$.
In the color system denoted with $I$ the pairs $Q_{1}\bar{Q}_{3}$ and $Q_{2}\bar{Q}_{4}$
form color singlets, and in the color system denoted with $I\hspace{-1.2pt}I$, the pairs $Q_{1}\bar{Q}_{4}$
and $Q_{2}\bar{Q}_{3}$ are color singlets. These two color states
are given by: 
\bea
|\mathcal{C}_{I}\rangle=\frac{1}{3} \delta_{c_1 \bar c_3} \delta_{c_2 \bar c_4} |c_1 c_2 \bar{c}_3\bar{c}_4\rangle
\non 
\\
|\mathcal{C}_{I\hspace{-1.2pt}I}\rangle=\frac{1}{3} \delta_{c_1 \bar c_4} \delta_{c_2 \bar c_3} |c_1 c_2 \bar{c}_3\bar{c}_4\rangle
\eea
However, these two states are not orthogonal, $\langle\mathcal{C}_{I}|\mathcal{C}_{I\hspace{-1.2pt}I}\rangle=\frac{1}{3}$.
For instance an orthogonal basis could be constructed by considering the antisymmetric $\cal A$
and symmetric $\cal S$ color combinations of these two states, 
\bea
|\mathcal{A}\rangle=\sqrt{\frac{3}{4}}\big(|\mathcal{C}_{I}\rangle-|\mathcal{C}_{I\hspace{-1.2pt}I}\rangle\big)
\non
\\
|\mathcal{S}\rangle=\sqrt{\frac{3}{8}}\big(|\mathcal{C}_{I}\rangle+|\mathcal{C}_{I\hspace{-1.2pt}I}\rangle\big)
\eea
and inversely,
\bea
|\mathcal{C}_{I}\rangle & = & \sqrt{\frac{2}{3}}|\mathcal{S}\rangle+\frac{1}{\sqrt{3}}|\mathcal{A}\rangle
\non 
\\
|\mathcal{C}_{I\hspace{-1.2pt}I}\rangle & = & \sqrt{\frac{2}{3}}|\mathcal{S}\rangle-\frac{1}{\sqrt{3}}|\mathcal{A}\rangle
\eea
It is useful to introduce
the contravariant basis to $|\mathcal{C}_{I}\rangle$
and $|\mathcal{C}_{I\hspace{-1.2pt}I}\rangle$. The color space metric matrix in this basis is, 
\begin{equation}
g=\left(\begin{array}{cc}
1 & 1/3\\
1/3 & 1
\end{array}\right)\label{eq:gdef}
\end{equation}
and it's inverse
\begin{equation}
g^{-1}=\left(\begin{array}{cc}
9/8 & -3/8\\
-3/8 & 9/8
\end{array}\right)
\end{equation}
Notice that $g$ is covariant while $g^{-1}$ is contravariant. Using
the metric matrices, we define the contravariant basis $|\mathcal{C}^{A}\rangle=g^{AB}|\mathcal{C}_{B}\rangle$,
with the property $\langle\mathcal{C}^{A}|\mathcal{C}_{B}\rangle=\delta_{B}^{A}$.
These contravariant basis is explicitly defined by 
\bea
|\mathcal{C}^{I}\rangle & = & \frac{9}{8}|\mathcal{C}_{I}\rangle-\frac{3}{8}|\mathcal{C}_{I\hspace{-1.2pt}I}\rangle \ ,
\non \\
|\mathcal{C}^{I\hspace{-1.2pt}I}\rangle & = & -\frac{3}{8}|\mathcal{C}_{I}\rangle+\frac{9}{8}|\mathcal{C}_{I\hspace{-1.2pt}I}\rangle \ .
\eea

\subsection{Completing the potential}

Since we have two independent color singlets, the potential must be given
by a $2\times2$ matrix. The lowest eigenvalue of the
matrix should correspond  to $V_{FF}$.

The corresponding eigenvectors $|u_{0}\rangle$ are the following,
\bi
\ie when $V_{FF}=V_{M\hspace{-1.2pt}M}$, the corresponding eigenvector is $|u_{0}\rangle=|\mathcal{C}_{I}\rangle$ ,
\ie $V_{FF}=V_{M\hspace{-1.2pt}M'}$ the corresponding eigenvector is $|u_{0}\rangle=|\mathcal{C}_{I\hspace{-1.2pt}I}\rangle$ ,
\ie when $V_{FF}=V_{T}$, we have $|u_{0}\rangle=|\mathcal{A}\rangle$.
\ei

The eigenvectors for the highest eigenvalue $|u_{1}\rangle$ must be orthogonal to $|u_{0}\rangle$ so that the potential matrix is hermitian. So, in the three cases, we have the following eigenvectors,
\bi
\ie
when $V_{FF}=V_{M\hspace{-1.2pt}M}$,
$|u_{1}\rangle=|\overline{C}_{I}\rangle=-\frac{1}{\sqrt{3}}|\mathcal{S}\rangle+\sqrt{\frac{2}{3}}|\mathcal{A}\rangle$
\ie
when $V_{FF}=V_{M\hspace{-1.2pt}M'}$,
$|u_{1}\rangle=|\overline{C}_{I\hspace{-1.2pt}I}\rangle=\frac{1}{\sqrt{3}}|\mathcal{S}\rangle+\sqrt{\frac{2}{3}}|\mathcal{A}\rangle$
\ie
when $V_{FF}=V_{T}$,
$|u_{1}\rangle=|\mathcal{S}\rangle$
\ei

To construct the potential matrix, we only need to know the highest
eigenvalue. In Refs. \cite{Bornyakov:2005kn,Cardoso:2012uka}, in the transition
region, there seems to be an exchange between the ground and the first
excited states. Namely, in region $A$ (where $V_{A}$ is the ground
state) close to the transition to region $B$ (where $V_{B}$ is the
ground state), the ground state's potential is $v_{0}=V_{A}$ and the
first excited state $v_{1}=V_{B}$. When we enter region $B$, then
we have $v_{0}=V_{B}$ and $v_{1}=V_{A}$. This way, the first excited
state should be the second lowest of the three potentials close to
the transitions. We assume, that this result is valid in general,
i e  
\bea
V_{FF}^1&=&\min \bigl[ \max(V_{M\hspace{-1.2pt}M},V_{M\hspace{-1.2pt}M'}),\max(V_{M\hspace{-1.2pt}M},V_{T}),
\non \\
&& \ \ \ \ \ \ 
\max(V_{M\hspace{-1.2pt}M'},V_{T}) \bigr] \ .
\eea

\subsection{Other Models \label{sec:othermodels}}

For comparison, we also use two other models.
One of them is similar to this color structure dependent
triple flip-flop model, but with the tetraquark sector removed from the potential. So, the ground state's potential
is just
\be
v_0 = \min(V_I, V_{II}) \ , 
\ee
and the excited state is,
\be
v_1 = \max(V_I, V_{II}) \ ,
\ee
with the corresponding eigenvectors, being constructed the same way as before. This potential is given in the matrix
form on the non-orthogonal basis formed by $|\mathcal{C}_I\rangle$ and $|\mathcal{C}_{II}\rangle$ by
\be
V = \left( 
	\begin{array}{cc}
	\frac{8}{9} V_I + \frac{1}{9} \min( V_I, V_{II} )  &  \frac{1}{3}\min(V_I,V_{II}) \\
    \frac{1}{3}\min(V_I,V_{II}) & \frac{8}{9} V_{II} + \frac{1}{9} \min( V_I, V_{II} ) \\
	\end{array}
\right) \ .
\ee

A third model we also compare with, is the colorless double Flip-flop. In this model, the potential is simply given by
\be
V = \min( V_I, V_{II} ) \ . 
\ee
This potential does not depend of the color degrees of freedom of the system and so is a non-physical one. However, this
model and it's extension to include the tetraquark sector have been used by several authors 
\cite{Lenz:1985jk,Oka:1984yx,Oka:1985vg}.   It can be interpreted as
the limit of a color structure dependent model, when the two eigenstates of the potential are degenerate ($v_0=v_1$).

\section{Our unitary technique to solve the meson-meson scattering and find tetraquark bound-states and resonances \label{sec:technique}}

\subsection{Meson-meson scattering equation}

Let us start with the microscopic Hamiltonian,
\begin{equation}
\hat{H}=\hat{T}_{Q}+\hat{V}_{Q} \ ,
\end{equation}
where the $Q$ subscript means that we are dealing with the kinetic
and potential energy of the quarks and not of the mesons. Folding
the operators in the color states $|\mathcal{C}_{I}\rangle$ and $|\mathcal{C}_{I\hspace{-1.2pt}I}\rangle$,
we get the Schr\"odinger equation:
\begin{equation}
\big(g_{AB}T_{Q}+V_{AB}^{Q}\big)\Psi^{B}=Eg_{AB}\Psi^{B} \ .
\end{equation}

We want to study meson-meson scattering because mesons and not quarks
are the asymptotic states of the theory. For this, we need the meson kinetic energy operator to show up
in the Schr\"odinger equation.
The mesons kinetic energy is given by
\begin{eqnarray*}
T_{M} & = & \left(\begin{array}{cc}
T_{Q}+V_{M\hspace{-1.2pt}M} & 0 \\
0 & T_{Q}+V_{M\hspace{-1.2pt}M'}
\end{array}\right) \ .
\end{eqnarray*}
W now need to opt for a decomposition of the potential and kinetic energies in the Hamiltonian.

However, due to the non-orthogonality of our color basis, it is not trivial to conveniently decompose the Hamiltonian. A first guess would be to consider the meson potential energy as,
\bea
V_{AB}^{M} &=& \left(\begin{array}{cc}
V_{I \, I} & V_{I \, I\hspace{-1.2pt}I}\\
V_{I\hspace{-1.2pt}I \, I} & V_{I\hspace{-1.2pt}I \, I\hspace{-1.2pt}I}
\end{array}\right)-\left(\begin{array}{cc}
1 & 1/3\\
1/3 & 1
\end{array}\right)\left(\begin{array}{cc}
V_{M\hspace{-1.2pt}M}& 0\\
0 & V_{M\hspace{-1.2pt}M'}
\end{array}\right)
\non \\ 
&=& \left(\begin{array}{cc}
V_{I \, I}-V_{M\hspace{-1.2pt}M} & V_{I \, I\hspace{-1.2pt}I}-\frac{1}{3}V_{M\hspace{-1.2pt}M'}\\
V_{I\hspace{-1.2pt}I \, I}-\frac{1}{3}V_{M\hspace{-1.2pt}M} & V_{I\hspace{-1.2pt}I \, I\hspace{-1.2pt}I}-V_{M\hspace{-1.2pt}M'}
\end{array}\right) \ ,
\eea
however this turns out to be a non-Hermitean operator, and clearly this is not a good decomposition of the Hamiltonian. 
Another possible decomposition would be to add the potentials $V_{M\hspace{-1.2pt}M}$ and
$V_{M\hspace{-1.2pt}M'}$ not to $T_{Q}\hat{1}$ but to $gT_{Q}$,
\begin{eqnarray*}
T_{M}' & = & \left(\begin{array}{cc}
T_{Q}+V_{M\hspace{-1.2pt}M} & \frac{1}{3}T_{Q}\\
\frac{1}{3}T_{Q} & T_{Q}+V_{M\hspace{-1.2pt}M'}
\end{array}\right) \ . 
\end{eqnarray*}
however we do not want $T_Q$ to appear in the equation.
We could as well define $T_{Q}=T_{M}^{A}-V_{M}^{A}$ which is convenient since
in one case the operator is applied to the two components of the wave-function.
However, numerically it could be difficult to verify
the equality $T_{M}^{I}-V_{M}^{I}=T_{M}^{I\hspace{-1.2pt}I}-V_{M}^{I\hspace{-1.2pt}I}$ . Moreover,
a simple reduced mass correction or the use of a relativistic quark
energy destroys this equality.

After considering several decompositions and testing them, we finally solve all these problems by considering the decomposition, 
\be
\hat{T}_{S}=\left(\begin{array}{cc}
\hat{T}_{M\hspace{-1.2pt}M} & \frac{\hat{T}_{M\hspace{-1.2pt}M}+\hat{T}_{M\hspace{-1.2pt}M'}}{6}\\
\frac{\hat{T}_{M\hspace{-1.2pt}M}+\hat{T}_{M\hspace{-1.2pt}M'}}{6} & \hat{T}_{M\hspace{-1.2pt}M'}
\end{array}\right) \ ,
\ee
and 
\be
\hat{V}_{S}=\left(\begin{array}{cc}
V_{I \, I}-V_{M\hspace{-1.2pt}M} & V_{I \, I\hspace{-1.2pt}I}-\frac{V_{M\hspace{-1.2pt}M}+V_{M\hspace{-1.2pt}M'}}{6}\\
V_{I \, I\hspace{-1.2pt}I}-\frac{V_{M\hspace{-1.2pt}M}+V_{M\hspace{-1.2pt}M'}}{6} & V_{I\hspace{-1.2pt}I \, I\hspace{-1.2pt}I}-V_{M\hspace{-1.2pt}M'}
\end{array}\right) \ .
\ee
Notice, both operators are now Hermitean. We have the Schr\"odinger equation,
\be
T_{AB}\Psi^{B}+V_{AB}\Psi^{B}=Eg_{AB}\Psi^{B} \ .
\label{schro_color}
\ee
The components $\Psi^{A}$ are expanded as,
\bea
\label{psiA_expansion}
\Psi^{I} &=& \sum_{i} \Phi^{I}_j (\boldsymbol{\rho}_{13},\boldsymbol{\rho}_{24}) \psi_{i}^{I}(\mathbf{r}_{1324})  \ ,
\non \\
\Psi^{II} &=& \sum_{i} \Phi^{II}_j (\boldsymbol{\rho}_{14},\boldsymbol{\rho}_{23}) \psi_{i}^{II}(\mathbf{r}_{1423})  \ ,
\eea
where we define the Jacobi coordinates,
\bea
\boldsymbol{\rho}_{13} &=& \mathbf{x}_1 - \mathbf{x}_3  \ ,
\non \\
\boldsymbol{\rho}_{24} &=& \mathbf{x}_2 - \mathbf{x}_4  \ ,
\non \\
\mathbf{r}_{1324} &=& \frac{m_1 \mathbf{x}_1 + m_3 \mathbf{x}_3}{m_1 + m_3} - \frac{m_2 \mathbf{x}_2 + m_4 \mathbf{x}_4}{m_2 + m_4}  \ ,
\non \\
\boldsymbol{\rho}_{14} &=& \mathbf{x}_1 - \mathbf{x}_4  \ ,
\non \\
\boldsymbol{\rho}_{23} &=& \mathbf{x}_2 - \mathbf{x}_3  \ ,
\non \\
\mathbf{r}_{1423} &=& \frac{m_1 \mathbf{x}_1 + m_4 \mathbf{x}_4}{m_1 + m_4} - \frac{m_2 \mathbf{x}_2 + m_3 \mathbf{x}_3}{m_2 + m_3}  \ .
\eea
The functions $\Phi^{A}_i$ are eigenfunctions of Hamiltonian of  the two non-interacting mesons,
\bea
(T_A + V_A) \Phi_i^A = E_{0i} \Phi_i^A \ .
\eea
They have the property,
\be
\lim_{\rho_1 \rightarrow \infty} \Phi_i^A(\boldsymbol{\rho}_1,\boldsymbol{\rho}_2) =
\lim_{\rho_2 \rightarrow \infty} \Phi_i^A(\boldsymbol{\rho}_1,\boldsymbol{\rho}_2) = 0 \ ,
\label{philimit}
\ee
because the quarks in each meson are confined.

Substituting Eq. (\ref{psiA_expansion}) into Eq. (\ref{schro_color}) and integrating the confined degrees of freedom,
(i. e. the $\boldsymbol{\rho}_i$ coordinates), we obtain the multichannel equation,
\be
\hat{T}_{\alpha\beta}\psi^{\beta}+\hat{V}_{\alpha\beta}\psi^{\beta}=E \hat{g}_{\alpha\beta}\psi^{\beta} \ .
\label{eq:schro_multichannel}
\ee
Here, we employ Greek indices to denote both the color structure and internal quantum numbers.

The operators $\hat{T}_{\alpha\beta}$, $\hat{V}_{\alpha\beta}$ and $\hat{g}_{\alpha\beta}$
are local between states with the same color structure but are non-local
between states with different color structures
Let us now analyze in detail the non-local operators.
Consider the non-diagonal term of the $g$ matrix, folding it with $\Psi^I$ and $\Psi^{II}$, 
\bea
\langle \Psi^{I*} | g_{I,II} | \Psi^{II} \rangle &=& \frac{1}{3} \sum_{ij}
\int d^3 \boldsymbol{\rho}_{13} d^3\boldsymbol{\rho}_{24} d^3 \mathbf{r}_{1324}
\non \\
&& \Phi^{I}_i ( \boldsymbol{\rho}_{13}, \boldsymbol{\rho}_{24} )^* \psi^I(\mathbf{r}_{1324})^*
\non \\
&& \Phi^{II}_j ( \boldsymbol{\rho}_{14}, \boldsymbol{\rho}_{23} ) \psi^{II}(\mathbf{r}_{1423}) \ .
\eea

Changing the integration variables to $\mathbf{r}_{1324}$, $\mathbf{r}_{1423}$ and
\be
\mathbf{r}_{1234} = \frac{m_1 \mathbf{x}_1 + m_2 \mathbf{x}_2}{m_1 + m_2} - \frac{m_3 \mathbf{x}_3 + m_4 \mathbf{x}_4}{m_3 + m_4} \ ,
\ee
in order to isolate functions of $\mathbf{r}_{1324}$ and $\mathbf{r}_{1423}$, the integral becomes
\be
\frac{1}{3} \int d^3 \mathbf{r}_{1324} d^3 \mathbf{r}_{1423} \,
\psi^I_i(\mathbf{r}_{1324})^* \gamma_{ij}( \mathbf{r}_{1324}, \mathbf{r}_{1423} ) \psi^{II}_j(\mathbf{r}_{1423}) \ .
\non
\ee
The function $\gamma_{ij}(\mathbf{r}_{1324}, \mathbf{r}_{1423})$ is given by,
\be
\gamma_{ij} = \Delta \int d^3 \mathbf{r}_{1234} \,  \Phi^{I}_i ( \boldsymbol{\rho}_{13}, \boldsymbol{\rho}_{24} )^* 
                       \Phi^{II}_j ( \boldsymbol{\rho}_{14}, \boldsymbol{\rho}_{23} ) \ ,
\label{eq:mesonmesonmetric}
\ee
where,
\be
\Delta = \Big( \frac{M_{12}M_{34}M_{13}M_{24}M_{14}M_{23}}{2 m_1 m_2 m_3 m_4 M^2} \Big)^3 \ ,
\ee
with $M_{ij} = m_i + m_j$ and $M = \sum_i m_i$.
In this way, knowing that $\hat{g}$ acts on a function as,
\be
\hat{g}_{\alpha\beta} \psi_\beta
= \int d^3 \mathbf{r}_{\beta} g_{\alpha\beta}(\mathbf{r}_\alpha,\mathbf{r}_\beta') \psi^{\beta}(\mathbf{r}_\beta') \ ,
\ee
we have,
\bea
g_{I\,i,I\,j}(\mathbf{r}_{1324},\mathbf{r}_{1324}') &=& \delta_{ij} \, \delta^3(\mathbf{r}_{1324} - \mathbf{r}_{1324}') \ ,
\non \\
g_{I\,i,II\,j}(\mathbf{r}_{1324},\mathbf{r}_{1423}') &=& \frac{1}{3} \gamma_{ij}(\mathbf{r}_{1324}, \mathbf{r}_{1423}') \ ,
\non \\
g_{II\,i,I\,j}(\mathbf{r}_{1423},\mathbf{r}_{1324}') &=& \frac{1}{3} \gamma_{ji}(\mathbf{r}_{1324}', \mathbf{r}_{1423})^* \ ,
\non \\
g_{II\,i,II\,j}(\mathbf{r}_{1324},\mathbf{r}_{1423}') &=& \delta_{ij} \, \delta^3(\mathbf{r}_{1423} - \mathbf{r}_{1423}') \ .
\eea

The potential $\hat{V}$ has a similar structure,
\bea
V_{A\,i,A\,j}(\mathbf{r}_{A},\mathbf{r}_{A}') &=& V^{AA}_{ij}(\mathbf{r}_{A})
\delta^3(\mathbf{r}_{A} - \mathbf{r}_{A}') \ ,
\non \\
V_{I\,i,II\,j}(\mathbf{r}_{1324},\mathbf{r}_{1423}') &=& v_{ij}(\mathbf{r}_{1324}, \mathbf{r}_{1423}') \ ,
\non \\
V_{II\,i,I\,j}(\mathbf{r}_{1324},\mathbf{r}_{1324}') &=& v_{ji}^*(\mathbf{r}_{1324}', \mathbf{r}_{1423}) \ ,
\eea
with
\bea
V^{I\, I}_{ij} &=& \int d^3 \boldsymbol{\rho}_{13} d^3 \boldsymbol{\rho}_{24}\,
\Phi^I_i(\boldsymbol{\rho}_{13},\boldsymbol{\rho}_{24})^* ( V_{I,I} - V_I ) \Phi^I_j(\boldsymbol{\rho}_{13},\boldsymbol{\rho}_{24}) \ ,
\non \\
V^{II\, II}_{ij} &=& \int d^3 \boldsymbol{\rho}_{14} d^3 \boldsymbol{\rho}_{23}\,
\Phi^I_i(\boldsymbol{\rho}_{14},\boldsymbol{\rho}_{23})^* ( V_{II,II} - V_{II} ) \Phi^I_j(\boldsymbol{\rho}_{14},\boldsymbol{\rho}_{23}) \ ,
\non \\
v_{ij} &=& \Delta \int d^3 \mathbf{r}_{1234} \,
\non \\
&& \Phi^I_i(\boldsymbol{\rho}_{13},\boldsymbol{\rho}_{24})^* ( V_{I,II} - \frac{V_{I}+V_{II}}{6} )
\Phi^{II}_j(\boldsymbol{\rho}_{14},\boldsymbol{\rho}_{23}) \ .
\label{eq:mesonmesonpot}
\eea
Note that, because of Eq. (\ref{philimit}), both $\gamma_{ij}$ and $v_{ij}$ have the property
\be
\lim_{r_1 \rightarrow \infty} \gamma_{ij}(\mathbf{r}_1,\mathbf{r}_2) = 
\lim_{r_2 \rightarrow \infty} \gamma_{ij}(\mathbf{r}_1,\mathbf{r}_2) = 0
\label{eq:gamma_limit}
\ee

The color structure preserving elements of $\hat{T}$ are just common kinetic energy
operators,
\be
\hat{T}_{Ai,Aj} = \delta_{ij} \big( E^A_{0i} - \frac{\hbar^2}{2 \mu^A_i} \nabla_A^2 \big)
\label{eq:opT_diag} \ ,
\ee
while the elements between states with different color structures are given by
\bea
\hat{T}_{Ii,IIj} &=& \frac{1}{6}
    \gamma_{ij}(\mathbf{r}_{1324},\mathbf{r}_{1423}) \big( E^{II}_{0i} - \frac{\hbar^2}{2 \mu^{II}_j} \nabla_{1423}^2 \big)
\non \\
    &&+ \frac{1}{6} \big( E^{I}_{0i} - \frac{\hbar^2}{2 \mu^{I}_j} \nabla_{1324}^2 \big)
    \gamma_{ij}(\mathbf{r}_{1324},\mathbf{r}_{1423}) \ .
\label{eq:opT_nondiag}
\eea

\subsection{Flux conservation}

Since $\hat{T}$, $\hat{V}$ and $\hat{g}$ are hermitian,we have the
continuity equation (for stationary states),
\begin{equation}
\Im[\psi^{\alpha*}\hat{T}_{\alpha\beta}\psi^{\beta}]=0 \ ,
\label{eq:continuity}
\end{equation}
or, in integral form,
\be
\sum_\alpha \Phi_\alpha = 0 \ ,
\ee
where
\be
\Phi_\alpha = \sum_\beta \int d^3 \mathbf{r}_\alpha \, 2 \Im[\psi^\alpha \hat{T}_{\alpha\beta} \psi^{\beta}]  \ ,
\ee
and the $2$ factor is just our convention.

Using the structure of $\hat{T}$ and the asymptotic behavior of $\gamma_{ij}$
given by Eqs. (\ref{eq:opT_diag}), (\ref{eq:opT_nondiag}) and (\ref{eq:gamma_limit}) we can easily prove that non-diagonal terms
on Eq. (\ref{eq:continuity}) do not contribute to the fluxes $\Phi_\alpha$. In his way, we get,
\be
\Phi_\alpha = \int dS \, \hat{\mathbf{n}}_\alpha \cdot \frac{\hbar}{2 i \mu_\alpha}
               ( \psi^{\alpha*} \nabla_\alpha \psi^{\alpha} - \psi^{\alpha} \nabla_\alpha \psi^{\alpha*}) \ .
\ee
Expanding each component $\psi^\alpha$ as,
\be
\psi^\alpha(\mathbf{r}) = \frac{u^\alpha(r)}{r} Y_{lm}(\theta_\alpha,\varphi_\alpha) \ ,
\ee
and taking the $r \rightarrow \infty$ limit we obtain,
\be
\Phi^\alpha = \lim_{r \rightarrow \infty} \frac{\hbar}{\mu_\alpha} \Im[u^{\alpha *} \frac{d u^\alpha}{dr}] \ .
\label{eq:flux_asymp}
\ee
Asymptotically $u^\alpha(r)$ behaves as,
\be
u^\alpha(r) \rightarrow A^\alpha \sqrt{\frac{\mu_\alpha}{k_\alpha}} \sin( k_\alpha r - \frac{l_\alpha \pi}{2} + \varphi_\alpha )
              + f_\alpha e^{i ( k_\alpha r - \frac{l_\alpha \pi}{2} )} \ .
\label{eq:asymp_behavior}
\ee
The term $\varphi_\alpha$ appears because $\hat{T}$ mixes the different channels.
Replacing Eq. \ref{eq:asymp_behavior} on Eq. \ref{eq:flux_asymp}, one gets,
\be
\sum_\alpha \sqrt{\frac{k_\alpha}{\mu_\alpha}} \Im[ A_\alpha f_\alpha e^{-i\varphi_\alpha} ] =
             \sum_\alpha \frac{k_\alpha}{\mu_\alpha} |f_\alpha|^2 \label{eq:opticth} \ .
\ee
This is the optical theorem for our system. Considering for instance the case of one channel it reads,
\be
\mbox{Im} \,t = |t|^2
\ee
with
\be
t = \sqrt{\frac{k}{\mu}} f e^{-i\varphi}
\ee

\subsection{T matrix}

To generalize this result, we first note that the wavefunction can be expanded as,
\be
u^\alpha(r) = \sum_i c_i u_i^\alpha(r) \label{eq:uexp1} \ ,
\ee
with,
\be
u_i^\alpha(r) \rightarrow
A_{i\alpha} \sqrt{\frac{\mu_\alpha}{k_\alpha}} \sin( k_\alpha r - \frac{l_\alpha \pi}{2} + \varphi_{i\alpha} )
              + f_{i\alpha} e^{i k_\alpha r- \frac{l_\alpha \pi}{2}} \label{eq:uexp2} \ .
\ee
Each $u_i^\alpha(r)$ can be decomposed in two parts,
\be
u_i^\alpha(r) = u_{0i}^\alpha(r) + v_i^\alpha(r) \ ,
\ee
where the function $u^\alpha_{0i}(r)$ comes from the eigenfunction $\psi^\alpha_0$ of $\hat{T}$, and has the asymptotic limit,
\be
u_{0i}^\alpha(r) \rightarrow
A_{i\alpha} \sqrt{\frac{\mu_\alpha}{k_\alpha}} \sin( k_\alpha r - \frac{l_\alpha \pi}{2} + \varphi_{i\alpha} ) \ ,
\ee
The $u^\alpha_{0i}$ can be chosen to form an orthonormal basis,
\be
\langle u_{0i} | u_{0j} \rangle = \frac{\pi}{2}\delta(E_i - E_j) \delta_{ij} \ ,
\ee
which imposes that the parameters $A_{i\alpha}$ and $\varphi_{i\alpha}$ must comply with the relation,
\be
\sum_\alpha A_{i\alpha} A_{j\alpha} \cos( \varphi_{i\alpha} - \varphi_{j\alpha} ) = \delta_{ij} \ .
\label{eq:orthoA}
\ee

Replacing Eqs. (\ref{eq:uexp1}) and (\ref{eq:uexp2}) in Eq. (\ref{eq:flux_asymp}), we are led to a generalization of Eq. (\ref{eq:opticth}),
\be
\sum_\alpha \sqrt{\frac{k_\alpha}{\mu_\alpha}}
\frac{A_{i\alpha} e^{-i\varphi_{i\alpha}} f_{j\alpha} - A_{j\alpha}  e^{-i\varphi_{j\alpha}} f_{i\alpha} }{2 i} =
             \sum_\alpha \frac{k_\alpha}{\mu_\alpha} f_{i\alpha}^* f_{j\alpha}  \ .
             \label{eq:genoptic}
\ee
We note that the left side of this equation has the form,
\be
\frac{T_{ij} - T_{ji}^*}{2i} \equiv \big( \frac{T-T^\dagger}{2i} \big)_{ij} \ ,
\ee
which suggests the definition of the $\mbox{T}$ matrix,
\be
T_{ij} = \sum_\alpha \sqrt{\frac{k_\alpha}{\mu_\alpha}} A_{i\alpha} e^{-i\varphi_{i\alpha}} f_{j\alpha} \ . \label{eq:tmat_def}
\ee

To verify Eq. (\ref{eq:tmat_def}), we use the relation,
\be
\sum_{k} A_{k\alpha} A_{k\beta} e^{i(\varphi_{k\alpha} - \varphi_{k\beta})} = \delta_{\alpha \beta} \ ,
\label{eq:completeA}
\ee
which can be proved from the completeness relation,
\be
\sum_n |\psi_n \rangle \langle \psi_n | = \hat{1} \ ,
\ee
of the eigenvectors of the kinetic energy $\hat{T}$ operator.
With Eq. (\ref{eq:completeA}(, we calculate $T^\dagger T$, and indeed we prove that it is equal to the right side of Eq. (\ref{eq:genoptic}) ,
\be
T^\dagger T = \sum_\alpha \frac{k_\alpha}{\mu_\alpha} f_{i\alpha}^* f_{j\alpha} \ .
\ee
Therefore, by the definition of $\mbox{T}$ and the previous relation, we can write relation \ref{eq:genoptic} as
\be
	\mbox{Im} \, \mbox{T} = \mbox{T}^\dagger \mbox{T}
\ee
which proves the unitarity of the $\mbox{S}$ matrix defined by $\mbox{S} = 1 + 2 i \mbox{T}$.

\subsection{Identical quarks and identical antiquarks}

So far, our framework is general for four-quark systems. We now specialize our study to a system of two identical quarks and two identical antiquarks, 

The total wavefunction must to be
antissymmetric under the quark-exchange $P_{12}$ and antiquark exchange $P_{34}$,
\bea
P_{12} \Psi &=& - \Psi \ ,
\non \\
P_{34} \Psi &=& - \Psi \ .
\eea
Given a generic wavefunction $\Psi$ we construct a completely antissymmetric on $\Psi_A$, by applying a projection operator
\be
\Psi_A = \mathcal{N} \big( 1 - P_{12} - P_{34} + P_{12} P_{34} \big) \Psi \label{antisymmetrize}
\ee

In this work, our Hamiltonian is spin-independent, and so we ignore spin effects in the dynamics of the system. With this approximation, the
spin part of the wavefunction factorizes, and we can neglect it's existence for everything except for the symmetry of
the wavefunction.
Now, consider the function,
\be
\Psi = \phi_\alpha(\boldsymbol{\rho}_{13}) \phi_\beta(\boldsymbol{\rho}_{24}) \psi(\mathbf{r}_{1324}) \mathcal{C}_{I} \Sigma
\ee
where $\Sigma$ is the spin part of the wavefunction and $\psi$ has parity
\be
\psi(-\mathbf{r}) = (-1)^{L_r} \psi(\mathbf{r})
\ee

We can make this function anti-symmetric by using Eq. (\ref{antisymmetrize}). In this case, we obtain the function,
\bea
\Psi_A &=& \mathcal{N} \Big[ \phi_\alpha(\boldsymbol{\rho}_{13}) \phi_\beta(\boldsymbol{\rho}_{24})
           \psi(\mathbf{r}_{1324}) \mathcal{C}_{I}
\non \\
&&         -(-1)^{1 + S_{12} + L_r}
           \phi_\alpha(\boldsymbol{\rho}_{23}) \phi_\beta(\boldsymbol{\rho}_{14}) \psi(\mathbf{r}_{1423}) \mathcal{C}_{II}
\non \\
&&         -(-1)^{1 + S_{34}}
           \phi_\alpha(\boldsymbol{\rho}_{14}) \phi_\beta(\boldsymbol{\rho}_{23}) \psi(\mathbf{r}_{1423}) \mathcal{C}_{II}
\non \\
&&         +(-1)^{S_{12}+S_{34}+L_r} \phi_\alpha(\boldsymbol{\rho}_{24}) \phi_\beta(\boldsymbol{\rho}_{13}) \psi(\mathbf{r}_{1324}) \mathcal{C}_{I}
\non \\
&&         \Big] \Sigma \ .
\label{eq:psiA}
\eea
We assume that $\Sigma$ is an eigenfunction of both $S_{12}$ and $S_{34}$. This is 
consistent with our approximation of neglecting all spin related interactions, which implies that all spin operators commute
trivially with the Hamiltonian.
Defining,
\bea
	s &\equiv& (-1)^{S_{12}+S_{34}+L_r}  \ ,
\non \\
	\xi &\equiv& (-1)^{S_{34}} \ ,
\eea
where $S_{12}$ and $S_{34}$ are respectively the spins of the $12$ and $34$ diquarks,
we are led to a simpler expression for $\Psi_A$,
\bea
	\Psi_A &= \mathcal{N}\big[& \Phi(\boldsymbol{\rho}_{13},\boldsymbol{\rho}_{24}) \psi(\mathbf{r}_{1324}) \mathcal{C}_I +
\non \\
          &&  \xi \Phi(\boldsymbol{\rho}_{14},\boldsymbol{\rho}_{23}) \psi(\mathbf{r}_{1423}) \mathcal{C}_{II} \big] \Sigma \ ,
\eea
with function $\Phi$ defined as,
\be
\Phi(\mathbf{x},\mathbf{y}) = \phi_\alpha(\mathbf{x}) \phi_\beta(\mathbf{y}) +
s\,\phi_\alpha(\mathbf{y})\phi_\beta(\mathbf{x}) \ ,
\ee
and having the property,
\be
\Phi(\mathbf{y},\mathbf{x}) = s\,\Phi(\mathbf{x},\mathbf{y}) \ .
\ee

Since both $S_{12}$ and $S_{34}$ are conserved in the non-dynamic spin approximation, we can diagonalize our Hamiltonian by
blocks with fixed values of these two operators. This is done by replacing Eq. (\ref{eq:psiA})
on Eq. (\ref{eq:schro_multichannel}). For this system, we obtain the scattering equation,
\bea
	&& \hat{T}_{\alpha\beta} \psi^\beta(\mathbf r) + V^D_{\alpha\beta} \psi^\beta(\mathbf r) +
\xi \int d^3 \mathbf{r}' v_{\alpha\beta}(\mathbf{r},\mathbf{r}') \psi^\beta(\mathbf{r}')
= \nonumber \\
&& E \big( \psi^\alpha
+ \xi \int d^3 \mathbf{r}' \gamma_{\alpha\beta}(\mathbf{r},\mathbf{r}') \psi^\beta(\mathbf{r}')
\big)  \ .
\label{eq:schro_exo}
\eea

Moreover since the total orbital angular momentum $L$ and parity are also conserved we can determine the possible "real"
quantum numbers of our system. 

For instance for $S_{12} = S_{34} = 0$, we only have $J = L$ states.  In this case, $\xi = 1$ and
$s = (-1)^{L_r}$.

For $S_{12} = S_{34} = 1$, we have $\xi = -1$ and still $s = (-1)^{L_r}$. The total angular momentum $J$ can range from
$|L - 2|$ to $L + 2$, as shown in Table \ref{tab:quantumn}.

\begin{table}[t!]
\begin{ruledtabular}
\begin{tabular}{|c|c|c|c|c|}
$S_{12}$ & $S_{34}$ & $\xi$ & $s(-1)^{L_r}$ & $J$ \tabularnewline
\hline 
0 & 0 & $+1$ & $+1$ & $L$ \\
0 & 1 & $-1$ & $-1$ & $|L-1|$ to $L+1$ \\
1 & 0 & $+1$ & $-1$ & $|L-1|$ to $L+1$ \\
1 & 1 & $-1$ & $+1$ & $|L-2|$ to $L+2$ \\
\end{tabular}
\end{ruledtabular}
\caption{ Quantum numbers of the $qq\bar{Q}\bar{Q}$ system }
\label{tab:quantumn}
\end{table}

\begin{table}[t!]
\begin{ruledtabular}
\begin{tabular}{|cc|cc|cc|}
$n_1$ & $L_1$ & $n_2$ & $L_2$ & $L_r$ & $s$ \tabularnewline
\hline 
0 & 0 & 0 & 0 & 0 & $+1$ \\
0 & 0 & 0 & 1 & 1 & $-1$ \\
0 & 0 & 1 & 0 & 0 & $+1$ \\
0 & 1 & 0 & 1 & 0 & $+1$ \\
\end{tabular}
\end{ruledtabular}
\caption{ Scattering channels used for the $qq\bar{Q}\bar{Q}$ system. }
\label{tab:channels}
\end{table}

\subsection{Numerical technique}

We consider the intra-mesonic potential to be of the funnel type, with no other corrections.
Moeover, we use a relativistic kinetic energy for the quarks in a meson,
\be
	\hat{T}_q = \sqrt{ m_1^2 + \mathbf{p}_1^2 } + \sqrt{ m_2^2 + \mathbf{p}_2^2 } \ .
\ee

As for the scattering problem, the kinetic energy of the mesons is considered to be non-relativistic,
but with the reduced mass obtained from the meson masses and not from the quark and antiquark masses,
\be
\mu_\alpha = \frac{M^\alpha_1 M^\alpha_2}{M^\alpha_1 + M^\alpha_2} \ ,
\ee
with $M_i^\alpha$ being the mass of the mesons.

Moreover, we neglect all other relativistic effects on our model, such as quark-antiquark annihilation, spin-spin and spin-orbit interactions. Besides this, we assume that we are well below the threshold for decay into a baryon-antibaryon system. In this way
we can neglected all the baryon-antibaryon channels, and consider just the meson-meson ones.

\subsubsection{Computation of meson-meson interaction}

To calculate the wave-functions of the mesons we use an harmonic oscillator
variational basis, where each function is given by,
\bea
\phi_{nlm}^{\beta}(r,\theta,\varphi) &=&
\sqrt{\frac{2n!\,\beta^{3}}{\Gamma(n+l+\frac{3}{2})}}\,(\beta r)^{l}\,\mbox{L}_{n}^{l+1/2}(\beta^{2}r^{2}) \, 
\non \\ &&
e^{-\frac{1}{2}\beta^{2}r^{2}}Y_{lm}(\theta,\varphi) \ .
\eea
$\mbox{L}_{n}^{l+1/2}$ are generalized Laguerre polynomials and $\beta$
is a variational parameter whose value we choose as $\beta=\left(\frac{4m\sigma}{3\sqrt{\pi}}\right)^{1/3}$.

The Meson Hamiltonian matrix elements are calculated in this basis and the matrix is diagonalized. With the resulting
eigenfunctions, we calculate the local $V_{ij}^{AA}$ and non-local $v_{ij}$ parts of the meson-meson potential,
defined by Eq. (\ref{eq:mesonmesonpot}) and the non-local parts of the metric matrix $\gamma_{ij}$
defined by Eq. (\ref{eq:mesonmesonmetric}).
These 7 (non-local) or 8 (local) dimensional integrals are calculated using Lebedev Quadrature in the angular coordinates and the Gauss-Laguerre Quadrature for the radial coordinates. We compute these multidimensional integrals using GPUs and the CUDA language.
The resulting functions depend on one radial coordinate in the local case and on two radial coordinates on the non-local case.
So, in both cases, the functions are evaluated on a 9-dimensional space.

\subsubsection{Meson-meson scattering}

We must solve Eq. \ref{eq:schro_multichannel}, to obtain the asymptotic behavior of the wavefunction and compute the T matrix.
This is done by  discretizing the radial scattering coordinates on a finite box.

We first solve the "free" equation 
\be
\hat{T}\Psi_{0}=Eg\Psi_{0} \ ,
\label{eq:free_schro}
\ee
This is not as simple to solve as it may seem, since we cannot solve this equation separately for each channel,
given the form of $T$ and $g$ matrices, which link all the channels. This is more cumbersome than what would happen in systems were there is only one type of quark combinations in mesons 
\cite{Bicudo:2010mv}, i. e. always $(q_1 \bar{q}_3) (q_2 \bar{q}_4)$ and never
$(q_1 \bar{q}_4) (q_2 \bar{q}_3)$.

For each energy, we generate different wavefunctions by varying the boundary conditions of the open channels.
The generated solutions are not orthogonal, however, and must be orthogonalized. Special care must be taken when doing
this, since, what we want is the asymptotic behavior of the functions. We can't consider the internal product of two
wavefunctions to be just given by the values on the finite box used for numerics,
\be
	\langle u | v \rangle = \sum_i u_i^* v_i \ ,
\ee
because, although the two functions $u$ and $v$ are orthogonal in the box, the continuations of them beyond the box
$\tilde{u}$ and $\tilde{v}$ aren't necessarily orthogonal,
\be
	\langle \tilde{u} | \tilde{v} \rangle \neq 0 \ .
\ee

Instead, we must utilize an inner product that takes into account the behavior of the wavefunction continued outside the box. To achieve this, we consider the inner product of two functions to be given by
\be
\langle u_i | u_j \rangle = \sum_\alpha A_{i\alpha} A_{j \alpha} \cos( \varphi_{i \alpha} - \varphi_{j \alpha} )
\label{eq:inner_asymp}
\ee
in accordance with Eq. (\ref{eq:orthoA}).
The parameters $A_{i\alpha}$ and $\varphi_{i\alpha}$ are given by Eq. (\ref{eq:asymp_behavior}) and are computed from the
value of the wavefunction components at the boundary of the box.

Having generated our basis of $N_{open}$ (the number of open channels) eigenfunctions of $\hat{T}$ for a given energy,
we orthogonalize the basis with the Gram-Schmidt procedure, applied  with the inner product of Eq. \ref{eq:inner_asymp}.

Then, for each function of the orthogonalized basis, we solve the scattering Eq. (\ref{eq:schro_multichannel})
considering $\Psi_{i}=\Psi_{0i}+\chi_{i}$. In this way the equation becomes
\be
(\hat{T}+\hat{V})\chi_{i}=Eg\chi_{i}-V\Psi_{0i} \, .
\ee
We solve this equation by discretizing it, using suitable boundary conditions for $\chi_i$. The values of $f_{i\alpha}$
are then calculated, from the behavior of $\chi_{i\alpha}$ at the boundary of the box. The $\mbox{T}$ matrix is evaluated utilizing Eq. (\ref{eq:tmat_def}).

\subsubsection{Finding bound states}

To find the bound states, of our system we don't need to calculate the $\mbox{T}$ matrix. We can directly  solve Eq. (\ref{eq:schro_multichannel}) and search for states with energy smaller than the first threshold.

The simplest method consists in discretizing Eq. (\ref{eq:schro_multichannel}) and solving it with Dirichlet boundary conditions.
However, this procedure only works as long as the spatial extent of the bound state is much smaller than the box on which we
are solving the equation. When this doesn't happen, the wavefunction can become highly distorted by the forced boundary conditions
and the energy can even be moved above the threshold, hiding the nature of the state.
In this work, as will be seen next, we can have bound states with a very small biding energy. For this states we would have to
use a very large box to obtain the energy and wavefunctions accurately. However, this is not convenient. Thus, instead of 
using Dirichlet boundary conditions, we consider boundary conditions that depend on the energy and match the expected behavior
of the wavefunctions at large distances.

With this method, we have to solve the matrix equation,
\be
\big[H + F(E)\big] \,u=Eg \, u \label{eq:bound_state} \ ,
\ee
where $F(E)$ is a matrix that fixes the boundary conditions.
This is not a simple eigenvalue problem as the matrix depends on the eigenvalue itself.

To solve this equation, we consider it as a root finding problem $\mbox{det} \big[H + F(E) - E g \big] = 0$
and use the Newton's method,
\be
x^{(n+1)} = x^{(n)} - \frac{f(x^{(n)})}{f'(x^{(n)})} \ ,
\ee
to solve it. 

Applying this method to Eq. (\ref{eq:bound_state}), we obtain the iteration procedure
\be
E^{(n+1)}=E^{(n)}-\frac{1}{\mbox{Tr}[(H+B(E)-Eg)^{-1}(B'(E)-g)]} \ ,
\ee
and we use it to calculate the bound-state energy. The wavefunction can then be calculated by solving Eq. (\ref{eq:bound_state})
as a simple linear system.

With this method, we find the bound states for the $x x \bar b \bar b$ detailed in Section \ref{sec:results}.

\subsubsection{Finding Resonances}

To find our resonance, we extend the $\mbox{T}$ matrix to the complex energy plane. 
Resonances correspond to complex poles of the $\mbox{S}$ (or $\mbox{T}$) matrix. So, to find resonances, we search for zeroes of the quantity
\be
y(E) = 1 / \mbox{Tr}[\mbox{T}] \, .
\label{eq:poleT}
\ee
Note that, on a pole, the $\mbox T$ matrix is divergent and so is it's
trace, therefore $y(E) = 0$.
Thus, to find a pole we just apply the Newton's method to Eq. (\ref{eq:poleT}).

\section{Results \label{sec:results}
}

\subsection{Bound states}

\begin{figure}[t!]
\begin{center}
\includegraphics[width=1.0\linewidth]{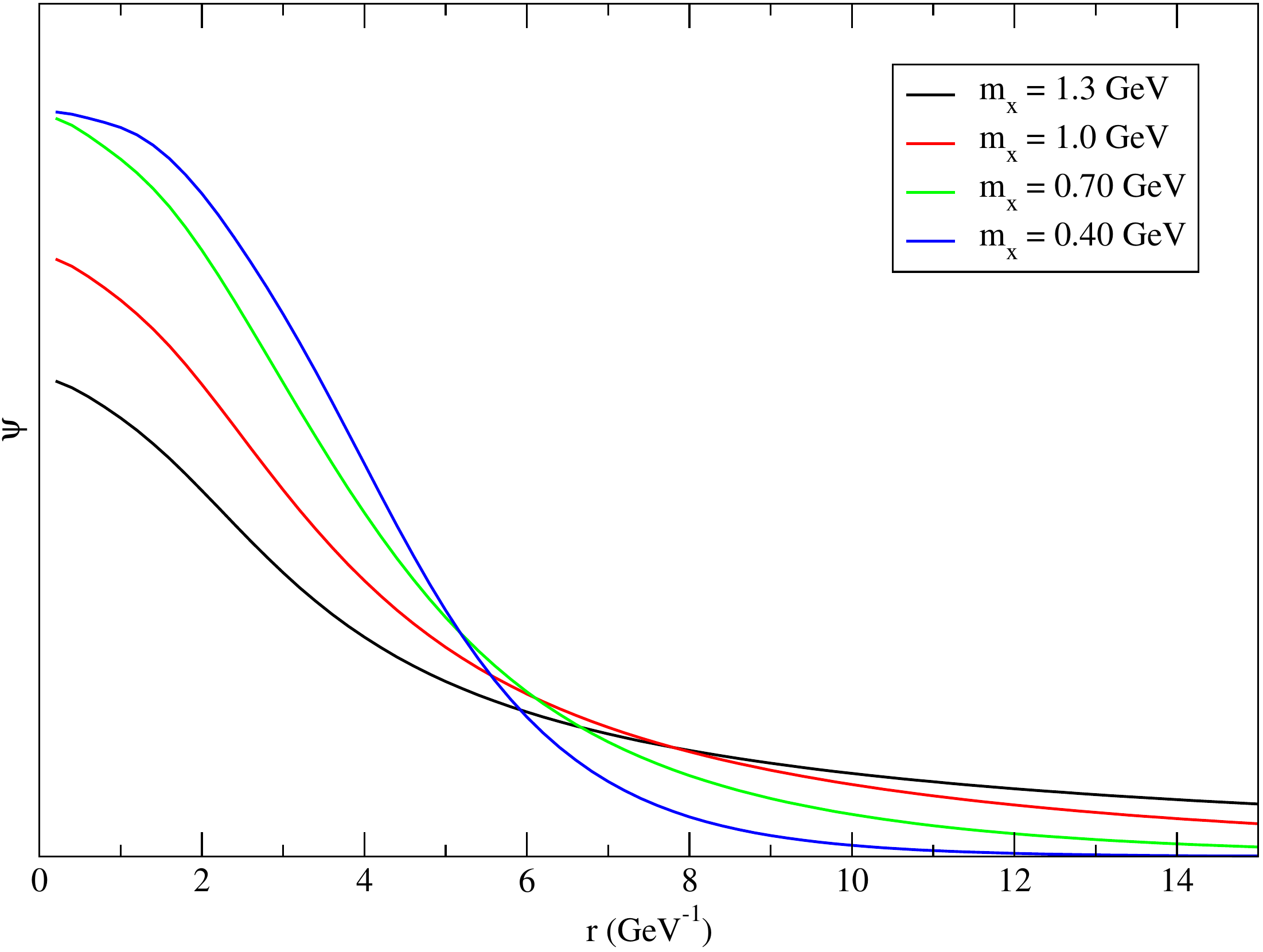}
\end{center}
\caption{Wavefunctions of the bound states we find in the $x x \bar b \bar b$ system.}
\label{fig:psi_bs}
\end{figure}

We apply the described method to find bound states in the the $qq\bar{b}\bar{b}$ system. We study different values of the light quark  mass, for angular momentum $L = 0$ and we are able to find bound states for quark masses ranging from $m_q = 0.4$ to $m_q = 1.3 \, \mbox{GeV}$. 
We are able to find bound states with $s(-1)^{L_r} = +1$, $\xi = +1$, listed  in Table \ref{tab:boundstates}. The boundstate is barely bound for $m_q = 1.3\, \mbox{GeV}$, the Charm mass, and becomes more strongly bound as $m_q$ decreases.
Wavefunctions for the first (see Table \ref{tab:channels}) dominant channel are given on Fig. \ref{fig:psi_bs}.

Now, comparing Table \ref{tab:quantumn}, these boundstates correspond to having $S_{12} = S_{34} = 0$, and hence to $L = 0$ and $J = 0$. 
If the lightest quarks are $u$ and $d$ quarks, we also have to consider the isospin symmetry, contributing an
additional $(-1)^{I_{12}+1}$ factor for the $P_{12}$ symmetry of the wavefunction. In this way, the previous results are unchanged
if $I_{12} = 1$. However if $I = 0$, the flavor wavefunction is anti-symmetric and so the spin wavefunction has to change it's
symmetry in order for the total wavefunction to be completely anti-symmetric. So, for $I_{12} = 0$, we have also $S_{12} = 1$,
and hence $S = 1$ and $J = 1$.

To summarize, we obtain tetraquark bound states on the $qq\bar{b}\bar{b}$ system, with quantum numbers $0^+$ for
s, c or b quarks, or light quarks with $I_{12} = 1$.
For light quarks with $I_{12} = 0$, we obtain bound states with quantum numbers $1^+$.

We also tried to find bound states for the $qq\bar{c}\bar{c}$ system, but we were unable to find them when the lightest quarks
have constituent masses equal or larger than the ones of light quarks $m_q \geq 400 \mbox{MeV}$.

\begin{table}[t!]
\begin{ruledtabular}
\begin{tabular}{|c|c|c|}
$m_q\mathrm{(GeV)}$ & $B\mathrm{(MeV)}$ & $R=\frac{1}{q_{0}}(\mbox{fm})$\tabularnewline
\hline 
1.30 & -0.02 & 18.8 \\
1.00 & -0.95 & 2.60 \\
0.70 & -7.91 & 0.92 \\
0.40 & -48.54 & 0.38 \\
\end{tabular}
\end{ruledtabular}
\caption{ Bound states in the system $qq\bar{b}\bar{b}$ with $L=0$ and $P=+$. We use $m_b = 4.7 GeV$.}
\label{tab:boundstates}
\end{table}

\subsection{Resonances}

With the effective meson-meson potential, we moreover compute the T Matrix and search for poles in the S matrix,
for the $q q \bar b \bar b$ system.

Searching for poles of the $S$ matrix in the complex energy plane for the $cc\bar{b}\bar{b}$ system with $L = 0$,
$P = +$ and $\xi = +1$, we find several resonances between all the four thresholds considered, as listed in Table
\ref{tab:resonances1} and showed in Fig. \ref{fig:argand_ccbb}.
The pole positions correspond to the two narrowest resonances of each threshold interval. The narrowest of all resonances appear between the opening of second and third channels (that $N_{open} = 2$ open channels).
Of the two narrowest resonances we find in this interval, one has a width of $16\, \mbox{MeV}$ and the other $36\,\mbox{MeV}$.
For the resonances found on the other two intervals, the narrowest of them all has a width of $140\,\mbox{MeV}$, much larger
than the ones found for $N_{open} = 2$. We remember that the second channel is the only of the four considered ones that has
an orbital angular momentum between the final state mesons. It is possible that the centrifugal barrier between the mesons is
what is slowing the decay of these tetraquark resonances, as discussed in Ref. \cite{Bicudo:2010mv}.

\begin{table}[t!]
\begin{ruledtabular}
\begin{tabular}{|c|c|c|c|}
$N_{open}$ & $E\mathrm{(GeV)}$ & $E_{n}$ & $E_{n+1}$ \tabularnewline
\hline 
1 & $12.6195 - 0.1248i$ & $12.5856$ & $12.974$ \\
  & $12.6797 - 0.1606i$ & & \\
2 & $12.9824 - 0.0078i$ & $12.974$ & $13.110$ \\
  & $12.9983 - 0.0179i$ &  & \\
3 & $13.1385 - 0.0701i$ & $13.110$ & $13.3624$ \\
  & $13.2121 - 0.0835i$ & & \\
\end{tabular}
\end{ruledtabular}
\caption{Resonances in the $0^{+}$ $cc\bar{b}\bar{b}$ system, with $\xi = +1$, for different numbers of open channels
$N_{open}$. $E_n$ and $E_{n+1}$ are the energies of the two thresholds. }
\label{tab:resonances1}
\end{table}

In Table \ref{tab:resonances2}, we list the two narrowest resonances between the second and third threshold, for different $m_q$ in the $0^{+}$ $cc\bar{b}\bar{b}$ system.
For all values of $m_q$, the narrowest of the two resonances has also the smallest $\mbox{Re}\, E$. It is interesting to note that, the width of these two resonances are inversely related. When one becomes larger the other decreases, with varying $m_q$.
Depending on the mass of the lightest quark, the width of these resonances can be smaller than $1\,\mbox{MeV}$ or larger than
$200\, \mbox{MeV}$. Both extreme cases happen for an unphysical mass $m_q = 700\, \mbox{MeV}$.

Broader resonances are also be found. In Fig. \ref{fig:argand_ccbb} we show a plot of the phase of $\Tr\ T$ for complex
energies with real parts between the second and third thresholds and negative imaginary parts. Four resonances are there
observed, all of them with widths smaller than $120\,\mbox{MeV}$. It is interesting to note the position of the four resonances
in the complex plane forming a straight line.

\begin{table}[t!]
\begin{ruledtabular}
\begin{tabular}{|c|c|c|c|}
$m_q\mathrm{(GeV)}$ & $E\mathrm{(GeV)}$ & $E_2$ & $E_3$ \tabularnewline
\hline 
1.30 & $12.9824 - 0.0078i$ & $12.974$ & $13.110$ \\
     & $12.9983 - 0.0179i$ &  & \\
1.00 & $12.4918 - 0.0048i$ & $12.4849$ & $12.6241$ \\
     & $12.5047 - 0.0192i$ & & \\
0.70 & $12.0329 - 2\times10^{-5}i$ & $12.0348$ & $12.1762$ \\
     & $12.0496 - 0.0215i$ & & \\
0.40 & $11.6163 - 0.0007i$ & $11.655$ & $11.796$ \\
     & $11.6661 - 0.0171i$ & & \\
\end{tabular}
\end{ruledtabular}
\caption{Resonances in the $0^{+}$ $qq\bar{b}\bar{b}$ system, with $\xi = +1$, with varying quark mass $m_q$.
$E_2$ and $E_3$ are the energy of the thresholds of the second and third channels. }
\label{tab:resonances2}
\end{table}

\begin{figure}[t!]
\begin{center}
\includegraphics[width=0.8\columnwidth, bb=60bp 200bp 550bp 592bp]{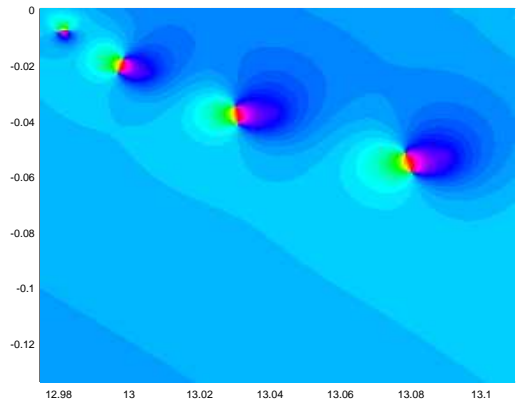}
\end{center}
\caption{Plot of $\Tr\ T$, between the second and third thresholds ($N_{open} = 2$), for the $cc\bar{b}\bar{b}$ system
with $L = 0$, $P = +$ and $\xi = +1$.
Note the existence of four resonances with $\Gamma / 2 < 60\, \mbox{MeV}$.
}
\label{fig:argand_ccbb}
\end{figure}

Doing the same study for $\xi = -1$, we also find several resonances. In Table \ref{tab:resonances3}, we compare
these resonances for the ones with $\xi = +1$ for $N_{open} = 2$ and $m_q = m_c$.
There, we can see that, in contrast to what happens for the bound states, not only
resonances exist for $\xi = -1$, but also have a behavior similar to that of $\xi = +1$. The real parts of the pole positions
are very close and the imaginary parts are of the same order of magnitude.
This result can be understood if we note that, what makes the results differ for the two values of $\xi$ is the presence of
the non-local potential. When we are above the threshold, the wave-function becomes oscillatory and the convolution of the
wave-function with the non-local potential, vanishes if the potential varies slowly. In this way, only the local
potential becomes important well above the threshold and, consequently,  the behavior of two resonances becomes similar.

\begin{table}[t!]
\begin{ruledtabular}
\begin{tabular}{|c|c|c|}
$N_{open}$ & \multicolumn{2}{c|}{$E(\mathrm{GeV})$} \tabularnewline
 & $\xi = +1$ & $\xi = -1$\tabularnewline
\hline
1 & $12.6195 - 0.1248i$ & $12.6069 - 0.0968i$ \\
  & $12.6797 - 0.1606i$ & $12.6659 - 0.1287i$ \\
2 & $12.9824 - 0.0078i$ & $12.9823 - 0.0106i$ \\
  & $12.9983 - 0.0179i$ & $13.0070 - 0.0256i$ \\
3 & $13.1385 - 0.0701i$ & $13.1272 - 0.0341i$ \\
  & $13.2121 - 0.0835i$ & $13.1532 - 0.0682i$ \\
\end{tabular}
\end{ruledtabular}
\caption{Comparison of pole positions found for $\xi = +1$ and $\xi = -1$ for the $0^{+}$ $qq\bar{b}\bar{b}$ system. }
\label{tab:resonances3}
\end{table}

For the $qq\bar{c}\bar{c}$ we are also able to find some resonances. The narrowest of them were between the opening of
the second and third channels, similarly to what happens on the $qq\bar{b}\bar{b}$ system. A study of these resonances
for $N_{open} = 2$ and varying $m_q$ is presented on Table \ref{tab:resonances4}. We find that the most stable of
the two resonances has a width that does not vary much with $m_q$, having always a value in the range of $35-41\,\mbox{MeV}$.
The width of the second in contrast increases monotonically from $76\,\mbox{MeV}$ to $108\,\mbox{MeV}$ as the mass $m_q$
decreases from $1.3\,\mbox{GeV}$ to $0.4\,\mbox{GeV}$.

\begin{table}[t!]
\begin{ruledtabular}
\begin{tabular}{|c|c|c|c|}
$m_{q}\mathrm{(GeV)}$ & $E\mathrm{(GeV)}$ & $E_2$ & $E_3$ \tabularnewline
\hline 
1.30 & $6.5036-0.0182i$ & $6.48644$ & $6.65375$ \\
     & $6.5427-0.0380i$ & & \\
1.00 & $5.9895-0.0194i$ & $5.97087$ & $6.14272$ \\
     & $6.0310-0.0413i$ & & \\
0.70 & $5.5113-0.0203i$ & $5.49134$ & $5.66693$ \\
     & $5.5547-0.0459i$ & & \\
0.40 & $5.0990-0.0184i$ & $5.07963$ & $5.25557$ \\
     & $5.1433-0.0541i$ &  & \\
\end{tabular}
\end{ruledtabular}
\caption{Resonances for the $0^{+}$ $qq\bar{c}\bar{c}$ system, with $\xi = +1$ and varying quark mass $m_q$. }
\label{tab:resonances4}
\end{table}

\begin{figure}[t!]
\begin{center}
\includegraphics[width=0.8\columnwidth, bb=60bp 200bp 550bp 592bp]{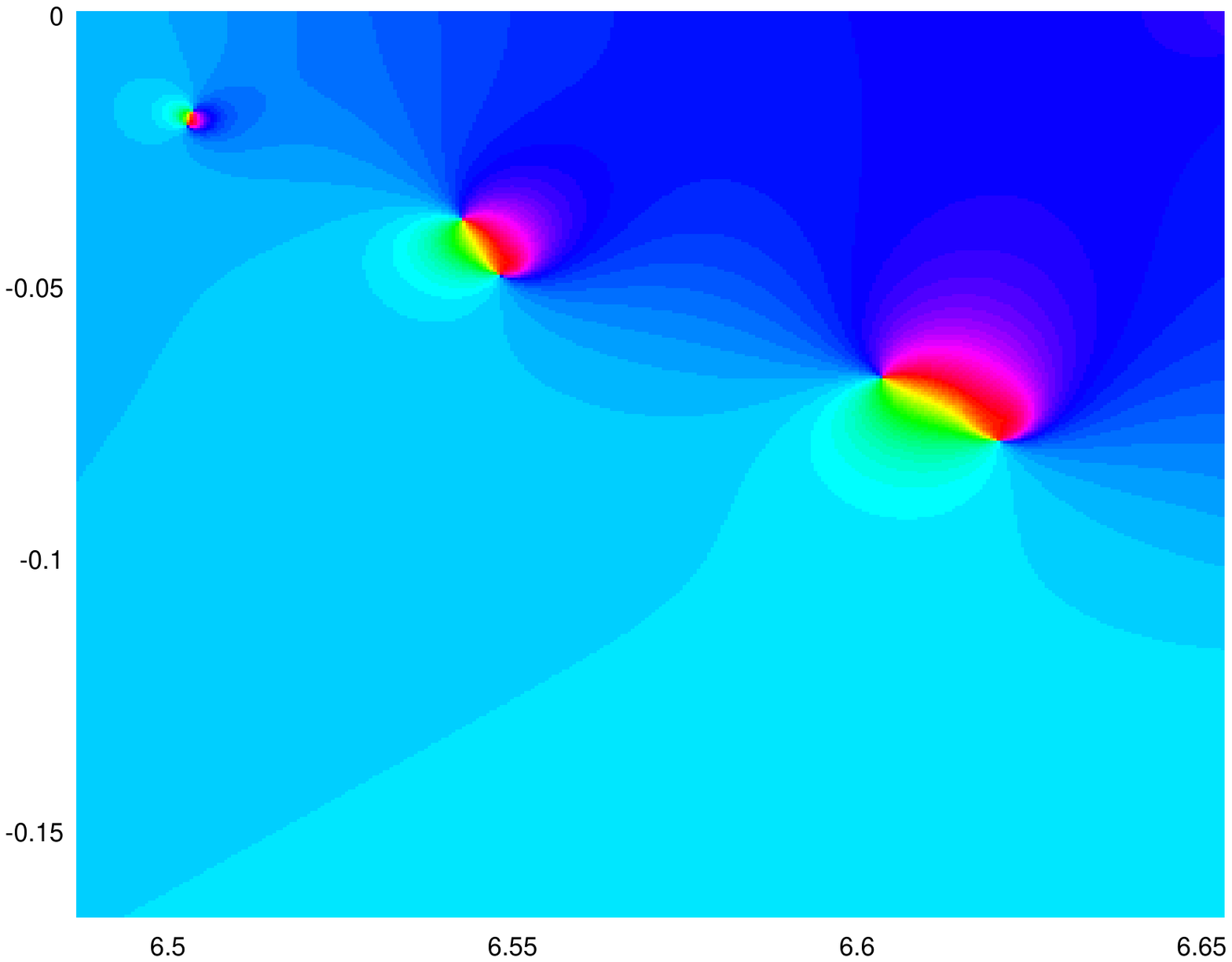}
\end{center}
\caption{Plot of $\Tr\ T$, between the second and third thresholds ($N_{open} = 2$), for the $cc\bar{c}\bar{c}$ system
with $L = 0$, $P = +$ and $\xi = +1$.
Note the existence of three resonances with $\Gamma / 2 < 70\, \mbox{MeV}$.
}
\label{fig:argand_ccbb}
\end{figure}

\subsection{Comparison with other models}

\begin{table}[t!]
\begin{ruledtabular}
\begin{tabular}{|c|c|c|c|}
$m_q\mathrm{(GeV)}$ & A & B & C \tabularnewline
\hline 
1.60 & -12.4 & - & - \\
1.30 & -13.3 & - & -0.02 \\
1.00 & -16.0 & -0.2 & -0.95 \\
0.70 & -25.1 & -5.7 & -7.91 \\
0.40 & -61.2 & -41.3 & -48.54 \\
\end{tabular}
\end{ruledtabular}
\caption{ Binding energies (in MeV) for the $0^+$ $qq\bar{b}\bar{b}$ system, $\xi=+1$, with three different models:
A - Color-independent simple flip-flop,
B - Color-dependent simple flip-flop,
C - Color-dependent triple flip-flop.
}
\label{tab:boundstatesABC}
\end{table}

We compare our model (C) with other two models for this system, defined in Section \ref{sec:othermodels}, namely one similar to our triple string flip-flop  potential, but without a tetraquark sector in
the potential( B) and the simple colorless flip-flop model (A).
The results of these three models are compared in Table \ref{tab:boundstatesABC}. As one sees there, when we remove the
tetraquark sector from our model, we still have bound states, for sufficiently small quark masses. For the color independent
simple flip-flop model, the binding is even bigger than in our model, even though the tetraquark sector is not present at all.

Therefore, we see that the presence of a tetraquark sector in the potential is not necessary for the existence of bound states
in the exotic tetraquark sector. This result has been observed by other authors, namely \cite{Lenz:1985jk,Oka:1984yx,Oka:1985vg} where bound states are found for a string flip-flop potential without any tetraquark configuration considered.

All these bound state results are for $s(-1)^{L_r} = +1$ and $\xi = +1$ 
(see Table \ref{tab:quantumn}),
which correspond to $S_{12} = S_{34} = 0$, and so $J^P=0^+$ for s, c or b quarks or, if $I_{12} = 1$, $S_{12} = 1$ and $S_{34} = 0$.
However, for $I_{12} = 0$, exchange symmetry imposes $S_{12} = 1$, which gives quantum numbers $1^+$.

\section{Discussion and conclusion \label{sec:conclu}}

In this work we briefly review recent experimental and lattice QCD results on tetraquarks. We extend the existing techniques to solve tetraquarks in quark models, fully unitarizing for the first time the triple string flip-flop potential to study boundstates and resonances in light-light-antiheavy-antiheavy systems $q q \bar Q \bar Q$. Consistently with the previous computations with simpler flip-flop potentials, we find several tetraquark boundstates and resonances \cite{Lenz:1985jk,Oka:1984yx,Oka:1985vg}. 

Comparing with recent lattice QCD results, \cite{Bicudo:2012qt,Bicudo:2015vta,Wagenbach:2014oxa,Scheunert:2015pqa},
which find boundstates but are not yet able to address resonances,
we find a qualitative dynamical agreement in the sense that biding is favored when the light quark $q$ gets lighter and the heavy antiquark $\bar Q$ gets heavier.

However, concerning the symmetry and quantum numbers, our results contradict the very recent lattice QCD results in Refs.  \cite{Bicudo:2012qt,Bicudo:2015vta,Wagenbach:2014oxa,Scheunert:2015pqa}.
In lattice QCD, attraction is only found in scalar isosinglet and vector
isotriplet channels, while here we only find bound states (and hence the maximum attraction) $S_{12} = 0$ and $I = 1$
(scalar isotriplet) and $S_{12} = 1$ and thus to $I = 0$ (vector isosinglet) channels.  Also note that the lattice results are consistent with the theoretical predictions of bound $q q \bar Q \bar Q$ systems with $\bar Q$ heavy enough
\cite{Ader:1981db,Ballot:1983iv,Heller:1986bt,Carlson:1987hh,Lipkin:1986dw,Brink:1998as,Gelman:2002wf,Vijande:2003ki,Janc:2004qn,Cohen:2006jg,Vijande:2007ix},
who imply the heavy anti-diquark $\bar Q \bar Q$ color wavefunction is a triplet 3. 

Notice we verified that the presence of the tetraquark sector in the flip-flop potential is not required for our tetraquark bound states.
Indeed, we note that the solutions with $\xi = +1$ are mostly color symmetric, while those with
$\xi = -1$ is mostly color anti-symmetric.
This means our bound states are mostly color symmetric, contrarily to what one would expect, but consistent with the fact that
removing the tetraquark sector from the potential does not destroy all the bound states.
As for the lattice results, they are indeed color anti-symmetric as one would expect. The attractive channels have the same
symmetry for spin and isospin as they are either scalar isosinglet ($S_{12} = 0$ and $I = 0$) or vector isotriplet
($S_{12} = 1$ and $I = 1$) and so, as the space symmetry is even, the color symmetry has to be negative for the total
wavefunction to be anti-symmetric.

Let us analyze the mechanism why we obtain attraction for the color symmetric case and repulsion in the color anti-symmetric case.
For the $qq\bar{Q}\bar{Q}$ system, if we consider only two channels (related by quark and antiquark exchange), the
scattering equation becomes
\bea
	&& \hat{T} \psi(\mathbf r) + V^D \psi(\mathbf r) +
\xi \int d^3 \mathbf{r}' v(\mathbf{r},\mathbf{r}') \psi(\mathbf{r}')
= \nonumber \\
&& E \big( \psi(\mathbf r)
+ \xi \int d^3 \mathbf{r}' \gamma(\mathbf{r},\mathbf{r}') \psi(\mathbf{r}') \ ,
\big)
\label{eq:schro_exo_1c}
\eea
calculated from Eq. \ref{eq:mesonmesonpot}. We see that it depends on $V_{I,II} - \frac{V_I + V_{II}}{6}$.
When the ground state is $v_0 = V_I$, we have $V_{I,II} = \frac{1}{3}V_I$ and so, this becomes $\frac{V_I - V_{II}}{6}$.
Since $V_I$ is the ground state, we have have $V_I < V_{II}$ and, therefore
\be
\frac{V_I - V_{II}}{6} < 0
\ee
Since the function $\Phi$ in the integrand is node-less (because it is the ground state),
\be
\Phi(\boldsymbol{\rho}_{13}, \boldsymbol{\rho}_{24})^* \Phi(\boldsymbol{\rho}_{14}, \boldsymbol{\rho}_{23}) > 0 \ .
\ee
Therefore, in this limit we have
\be
v(\mathbf{r},\mathbf{r}') < 0
\ee
for $r' \rightarrow \infty$ and fixed $r$. The same result occurs if we exchange $r$ and $r'$.
Then, in Eq. (\ref{eq:schro_exo_1c}) the non-local potential will be mostly attractive for $\xi = +1$
and mostly repulsive for $\xi = -1$. This confirms our results.

To conclude we develop fully unitarized techniques to study tetraquarks with quark models, and to search for boundstates and resonances.  Asymptotically, the four quark system with a string flip-flop potential reduces to coupled two-body meson-meson systems, with non-local potentials that vanish at long distances, thus solving the Van der Waals problem. However, we find that the string flip-flop potential still remains attractive enough to produce $qq\bar{Q}\bar{Q}$ boundstates with quantum numbers not observed in recent lattice QCD computations. 

It should be noted that different solutions to remove the excessive attraction exist \cite{Miller:1987wv}. For instance, attraction may change if the non-local potentials
change sign in the region where $r_1 \sim r_2$, and the presence of the $g_{\alpha\beta}$ operator cancels in part this effect.
We expect that our results will motivate future studies of different solutions to this excessive attraction.

Marco Cardoso is supported by FCT under the contract SFRH/BPD/73140/2010.


\bibliographystyle{apsrev4-1}
\bibliography{xxBB.bib}

\end{document}